\documentclass[ALICE,manyauthors]{cernphprep}
\usepackage[comma,square,numbers,sort&compress]{natbib}
\usepackage{hyperref}
\usepackage{booktabs}
\usepackage{lineno}
\usepackage{color}
\usepackage{cleveref} 
\newcolumntype{Y}{>{\centering\arraybackslash}X}
\usepackage[T1]{fontenc}
\usepackage{orcidlink}

\RequirePackage[normalem]{ulem} 

\begin{document}
%

\newcommand{\red}{\textcolor{red}}
\newcommand{\blue}{\textcolor{blue}}

\newcommand{\orho}{\ensuremath{1-\rho}\xspace}
\newcommand{\Rpp}{\ensuremath{R_{\mathrm{pp}}}\xspace}

\newcommand{\ST}{\ensuremath{S_{\rm T}}\xspace}
\newcommand{\RT}{\ensuremath{R_{\rm T}}\xspace}
\newcommand{\Nm}{\ensuremath{N_{\mathrm{m}}}\xspace}
\newcommand{\Nt}{\ensuremath{N_{\mathrm{t}}}\xspace}
\newcommand{\NT}{\ensuremath{N_{\mathrm{T}}}\xspace}
\newcommand{\mNT}{\ensuremath{\langle N_{\mathrm{T}}} \rangle \xspace}
\newcommand{\Smt}{\ensuremath{S_{\mathrm{mt}}}\xspace}

\newcommand{\pipm}{\ensuremath{\pi^{\pm}}\xspace}
\newcommand{\kpm}{\ensuremath{\mathrm{K}^{\pm}}\xspace}
\newcommand{\ppm}{\ensuremath{(\overline{\mathrm{p}})\mathrm{p}}\xspace}


\newcommand{\RMp}{\ensuremath{P(N_{\mathrm{T,m}}|N_{\mathrm{T,t}})}\xspace}
\newcommand{\NTc}{\ensuremath{N_{\mathrm{T}}}\xspace}
\newcommand{\NTt}{\ensuremath{N_{\mathrm{T,t}}}\xspace}
\newcommand{\NTm}{\ensuremath{N_{\mathrm{T,m}}}\xspace}
\newcommand{\UM}{\ensuremath{M_{\mathrm{tm}}}\xspace}

\newcommand{\UnfDis}{\ensuremath{Y(N_{\mathrm{T,t}}})\xspace}
\newcommand{\RawDis}{\ensuremath{Y(N_{\mathrm{T,m}}})\xspace}
\newcommand{\Mone}{\ensuremath{\mathrm{M1}_{\mathrm{tm}}}\xspace}
\newcommand{\Mtwo}{\ensuremath{\mathrm{M2}_{\mathrm{tm}}(p_{\mathrm{T}})}\xspace}
\newcommand{\Prior}{\ensuremath{P_{0}(\NTm)}\xspace}
\newcommand{\UpdatedPrior}{\ensuremath{\widehat{P}(\NTt)}\xspace}
\newcommand{\RawYield}{\ensuremath{\mathrm{d}Y(N_{\mathrm{T,m}})/\mathrm{d}p_{\mathrm{T}}}}
\newcommand{\UnfYield}{\ensuremath{\mathrm{d}Y(N_{\mathrm{T,t}})/\mathrm{d}p_{\mathrm{T}}}}

\newcommand{\dndetainline}{\ensuremath{\langle\text{d}N_{\rm{ch}}/\text{d}\eta\rangle}\xspace}
\newcommand{\dndyinline}{\ensuremath{\text{d}N/\text{d}y}\xspace}
\newcommand{\dndydptinline}{\ensuremath{\text{d}^2N/\text{d}\ptt \text{d}y}\xspace}

\newcommand{\pp}           {pp\xspace}
\newcommand{\ppbar}        {\mbox{$\mathrm {p\overline{p}}$}\xspace}
\newcommand{\XeXe}         {\mbox{Xe--Xe}\xspace}
\newcommand{\PbPb}         {\mbox{Pb--Pb}\xspace}
\newcommand{\pA}           {\mbox{pA}\xspace}
\newcommand{\pPb}          {\mbox{p--Pb}\xspace}
\newcommand{\AuAu}         {\mbox{Au--Au}\xspace}
\newcommand{\dAu}          {\mbox{d--Au}\xspace}

\newcommand{\s}            {\ensuremath{\sqrt{s}}\xspace}
\newcommand{\snn}          {\ensuremath{\sqrt{s_{\mathrm{NN}}}}\xspace}
\newcommand{\pt}           {\ensuremath{p_{\rm T}}\xspace}
\newcommand{\mt}           {\ensuremath{m_{\mathrm{T}}}\xspace}
\newcommand{\ptleading}    {\ensuremath{p_{\rm T}^{\rm{leading}}}\xspace}
\newcommand{\meanpt}       {\ensuremath{\langle p_{\rm T}\rangle}\xspace}
\newcommand{\ycms}         {\ensuremath{y_{\rm CMS}}\xspace}
\newcommand{\ylab}         {\ensuremath{y_{\rm lab}}\xspace}
\newcommand{\etarange}[1]  {\mbox{$\left | \eta \right |~<~#1$}}
\newcommand{\yrange}[1]    {\mbox{$\left | y \right |~<~#1$}}
\newcommand{\dndy}         {\ensuremath{\mathrm{d}N/\mathrm{d}y}\xspace}
\newcommand{\dndeta}       {\ensuremath{\mathrm{d}N_\mathrm{ch}/\mathrm{d}\eta}\xspace}
\newcommand{\avdndeta}     {\ensuremath{\langle\dndeta\rangle}\xspace}
\newcommand{\dNdy}         {\ensuremath{\mathrm{d}N_\mathrm{ch}/\mathrm{d}y}\xspace}
\newcommand{\Npart}        {\ensuremath{N_\mathrm{part}}\xspace}
\newcommand{\Ncoll}        {\ensuremath{N_\mathrm{coll}}\xspace}
\newcommand{\dEdx}         {\ensuremath{\textrm{d}E/\textrm{d}x}\xspace}
\newcommand{\meandEdx}{\ensuremath{\langle\textrm{d}E/\textrm{d}x}\rangle\xspace}

\newcommand{\RpPb}         {\ensuremath{R_{\rm pPb}}\xspace}
\newcommand{\Qpp}         {\ensuremath{Q_{\rm pp}}\xspace}

\newcommand{\nineH}        {$\sqrt{s}~=~0.9$~Te\kern-.1emV\xspace}
\newcommand{\seven}        {$\sqrt{s}~=~7$~Te\kern-.1emV\xspace}
\newcommand{\twoH}         {$\sqrt{s}~=~0.2$~Te\kern-.1emV\xspace}
\newcommand{\twosevensix}  {$\sqrt{s}~=~2.76$~Te\kern-.1emV\xspace}
\newcommand{\five}         {$\sqrt{s}~=~5.02$~Te\kern-.1emV\xspace}
\newcommand{\thirteen}    {$\sqrt{s}~=~13$~Te\kern-.1emV\xspace}
\newcommand{\twosevensixnn}{$\sqrt{s_{\mathrm{NN}}}~=~2.76$~Te\kern-.1emV\xspace}
\newcommand{\fivenn}       {$\sqrt{s_{\mathrm{NN}}}~=~5.02$~Te\kern-.1emV\xspace}
\newcommand{\LT}           {L{\'e}vy-Tsallis\xspace}
\newcommand{\GeVc}         {Ge\kern-.1emV/$c$\xspace}
\newcommand{\MeVc}         {Me\kern-.1emV/$c$\xspace}
\newcommand{\TeV}          {Te\kern-.1emV\xspace}
\newcommand{\GeV}          {Ge\kern-.1emV\xspace}
\newcommand{\MeV}          {Me\kern-.1emV\xspace}
\newcommand{\GeVmass}      {Ge\kern-.2emV/$c^2$\xspace}
\newcommand{\MeVmass}      {Me\kern-.2emV/$c^2$\xspace}
\newcommand{\lumi}         {\ensuremath{\mathcal{L}}\xspace}
\newcommand{\gevc}[1]      {\ensuremath{#1 \text{\,GeV/$c$}}\xspace}
\newcommand{\mevc}[1]      {\ensuremath{#1 \text{\,MeV/$c$}}\xspace}
\newcommand{\sppt}[1]      {\ensuremath{\sqrt{s} = #1 \text{\,TeV}}\xspace}

\newcommand{\ITS}          {\rm{ITS}\xspace}
\newcommand{\TOF}          {\rm{TOF}\xspace}
\newcommand{\ZDC}          {\rm{ZDC}\xspace}
\newcommand{\ZDCs}         {\rm{ZDCs}\xspace}
\newcommand{\ZNA}          {\rm{ZNA}\xspace}
\newcommand{\ZNC}          {\rm{ZNC}\xspace}
\newcommand{\SPD}          {\rm{SPD}\xspace}
\newcommand{\SDD}          {\rm{SDD}\xspace}
\newcommand{\SSD}          {\rm{SSD}\xspace}
\newcommand{\TPC}          {\rm{TPC}\xspace}
\newcommand{\TRD}          {\rm{TRD}\xspace}
\newcommand{\VZERO}        {\rm{V0}\xspace}
\newcommand{\VZEROA}       {\rm{V0A}\xspace}
\newcommand{\VZEROC}       {\rm{V0C}\xspace}
\newcommand{\Vdecay} 	   {\ensuremath{V^{0}}\xspace}

\newcommand{\hadrons}      {\ensuremath{\mathrm{h}^{\pm}}\xspace}
\newcommand{\pion}         {\ensuremath{\pi}\xspace}
\newcommand{\kaon}         {\ensuremath{\textrm{K}}\xspace}
\newcommand{\pr}           {\ensuremath{\textrm{p}}\xspace}
\newcommand{\prx}          {\ensuremath{\mathrm{p}(\overline{\mathrm{p}}})\xspace}
\newcommand{\ee}           {\ensuremath{e^{+}e^{-}}} 
\newcommand{\pip}          {\ensuremath{\pi^{+}}\xspace}
\newcommand{\pim}          {\ensuremath{\pi^{-}}\xspace}
\newcommand{\kapm}         {\ensuremath{\mathrm{K}^{\pm}}}
\newcommand{\kap}          {\ensuremath{\rm{K}^{+}}\xspace}
\newcommand{\kam}          {\ensuremath{\rm{K}^{-}}\xspace}
\newcommand{\pbar}         {\ensuremath{\rm\overline{p}}\xspace}
\newcommand{\kzero}        {\ensuremath{{\rm K}^{0}_{\rm{S}}}\xspace}
\newcommand{\lmb}          {\ensuremath{\Lambda}\xspace}
\newcommand{\almb}         {\ensuremath{\overline{\Lambda}}\xspace}
\newcommand{\Om}           {\ensuremath{\Omega^-}\xspace}
\newcommand{\Mo}           {\ensuremath{\overline{\Omega}^+}\xspace}
\newcommand{\X}            {\ensuremath{\Xi^-}\xspace}
\newcommand{\Ix}           {\ensuremath{\overline{\Xi}^+}\xspace}
\newcommand{\Xis}          {\ensuremath{\Xi^{\pm}}\xspace}
\newcommand{\Oms}          {\ensuremath{\Omega^{\pm}}\xspace}
\newcommand{\degree}       {\ensuremath{^{\rm o}}\xspace}

\newcommand{\ktopi}        {\ensuremath{\mathrm{K}/\pi}\xspace}
\newcommand{\ptopi}        {\ensuremath{\mathrm{p}/\pi}\xspace}

\newcommand{\nsigma}       {\ensuremath{\mathrm{n}_{\sigma}}\xspace}

\newcommand{\py}{PYTHIA~8\xspace}
\newcommand{\ep}{EPOS~LHC\xspace}
\newcommand{\gea}{GEANT~3\xspace}

\begin{titlepage}
\PHyear{2024}       
\PHnumber{205}      
\PHdate{22 July}  

\title{Particle production as a function of charged-particle flattenicity in pp collisions at $\mathbf{\sqrt{{\textit s}}=13}$~\textbf{TeV}}
\ShortTitle{Particle production as a function of charged-particle flattenicity in pp collisions}   

\Collaboration{ALICE Collaboration\thanks{See Appendix~\ref{app:collab} for the list of collaboration members}}
\ShortAuthor{ALICE Collaboration} 


\begin{abstract}
This paper reports the first measurement of the transverse momentum ($p_{\mathrm{T}}$) spectra of primary charged pions, kaons, (anti)protons, and unidentified particles as a function of the charged-particle flattenicity in pp collisions at $\sqrt{s}=13~\mathrm{TeV}$. Flattenicity is a novel event shape observable that is measured in the pseudorapidity intervals covered by the V0 detector, $2.8<\eta<5.1$ and $-3.7<\eta<-1.7$. According to QCD-inspired phenomenological models, it shows sensitivity to multiparton interactions and is less affected by biases toward larger $p_{\mathrm{T}}$ due to local multiplicity fluctuations in the V0 acceptance than multiplicity. The analysis is performed in minimum-bias (MB) as well as in high-multiplicity events up to $p_{\mathrm{T}}=\gevc{20}$. The event selection requires at least one charged particle produced in the pseudorapidity interval $|\eta|<1$. The measured $p_{\mathrm{T}}$ distributions, average $p_{\mathrm{T}}$, kaon-to-pion and proton-to-pion particle ratios, presented in this paper, are compared to model calculations using PYTHIA~8 based on color strings and EPOS~LHC. The modification of the $p_{\mathrm{T}}$-spectral shapes in low-flattenicity events that have large event activity with respect to those measured in MB events develops a pronounced peak at intermediate $p_{\mathrm{T}}$ ($2<p_{\mathrm{T}}<8$\,GeV/$c$), and approaches the vicinity of unity at higher $p_{\mathrm{T}}$. The results are qualitatively described by PYTHIA, and they show different behavior than those measured as a function of charged-particle multiplicity based on the V0M estimator.
\end{abstract}
\end{titlepage}

\setcounter{page}{2} 


\section{Introduction} 
\label{sec:introduction}

In proton--proton (pp) collisions at the LHC energies, hard parton--parton scatterings with momentum transfer above several\,GeV/$c$ produce high transverse momentum (\pt) particles that can be described by perturbative quantum chromodynamics (pQCD). Additional parton--parton scatterings that are not part of the main hard process constitute the underlying event (UE), which is modeled using phenomenological approaches~\cite{Diehl:2011yj,Blok:2011bu}. At LHC energies, the large parton densities result in a significant probability of more than one partonic interaction in a single pp collision~\cite{PhysRevD.36.2019}, a phenomenon known as multiparton interaction (MPI) that is supported by data~\cite{ALICE:2013tla, ALICE:2012cor}. In MPI-based models, pp collisions with high charged-particle multiplicities are dominantly those with a larger-than-average number of MPIs. The properties of the hadronic final state are sensitive to the interplay between the final states of several parton-parton interactions, the modeling of MPI, and non-perturbative final-state effects such as color reconnection (CR) implemented in \py~\cite{PhysRevD.36.2019,Sjostrand:2014zea}. For example, CR in pp collisions containing a large amount of MPI creates a strong correlation between the average transverse momentum of the produced particles and the charged particle multiplicity~\cite{ATLAS:2010jvh}. The strength of this correlation is mass dependent, and therefore, reminiscent of radial flow effects in heavy-ion collisions~\cite{OrtizVelasquez:2013ofg}.

Recent measurements in small collision systems such as high-multiplicity (HM) pp and p--Pb collisions at the LHC have revealed several effects that are qualitatively similar to the ones observed in heavy-ion collisions. Such phenomena include collective flow~\cite{CMS:2010ifv,ATLAS:2015hzw,CMS:2015fgy,ALICE:2012eyl,ATLAS:2016yzd,CMS:2013jlh,ALICE:2013snk,CMS:2014und,CMS:2012qk,ATLAS:2012cix,LHCb:2015coe,ALICE:2020nkc,CMS:2016zzh,CMS:2016fnw,ALICE:2019zfl} and the enhanced production of strange hadrons with respect to the charged-pion yield~\cite{ALICE:2018pal,ALICE:2016fzo,ALICE:2019avo,ALICE:2013wgn,ALICE:2015mpp}. Despite the large amount of soft-QCD results on collectivity, the origin of these phenomena in small systems is not yet fully understood. For example, experimental searches for jet modifications due to the presence of a medium in small collision systems have not been successful within current experimental precision, though its effects are expected to be small~\cite{ALICE:2023plt,Nagle:2018nvi,ALICE:2015umm,ALICE:2016faw,ALICE:2017svf,ALICE:2022qxg,ALICE:2022fnb,ALICE:2023csm}. Moreover, recent results from ALICE suggest that the measured ridge yields in low-multiplicity pp collisions are nonzero and substantially larger than the limits set in $\mathrm{e}^{+}\mathrm{e}^{-}$ annihilation~\cite{ALICE:2023ulm}. ATLAS also observed significant nonzero values of the second- and third-order flow coefficients measured in photonuclear ultraperipheral Pb--Pb collisions~\cite{ATLAS:2021jhn}. Thus, the existing measurements do not yet provide an answer to the important question of whether the origin of collectivity in small systems is attributed to the formation of a strongly-interacting quark--gluon plasma (QGP), or if it originates from different physical mechanisms. Several theoretical approaches have been suggested to explain the QGP-like effects in small collision systems. For example, the \py~\cite{Sjostrand:2014zea} model can qualitatively describe some of the observed features by incorporating new phenomenological final-state prehadronization mechanisms, such as rope hadronization~\cite{Bierlich:2014xba}, string shoving~\cite{Bierlich:2016vgw}, and MPI together with the CR mechanisms~\cite{Sjostrand:2014zea,OrtizVelasquez:2013ofg}.

The production of (un)identified charged hadrons as a function of multiplicity has been studied to understand the origin of the collective-like effects observed in pp and p--Pb collisions. Measurements of the inclusive charged particle production as a function of multiplicity indicate a stronger-than-linear increase of the high-$p_{\rm T}$ particle yields with increasing multiplicity relative to the yield in minimum-bias (MB) pp collisions~\cite{ALICE:2019dfi}, which is a consequence of an autocorrelation bias. To minimize such biases, the event classification has also been performed using charged-particle multiplicity measurements at forward pseudorapidity, i.e.\ in a different pseudorapidity interval than the one in which the observable of interest is measured~\cite{ALICE:2018pal}. However, this event selection approach is still sensitive to biases from local multiplicity fluctuations originating from jets that in turn enhance the high-$p_{\rm T}$ particle production, affecting the search for medium-induced jet modification in small systems~\cite{ALICE:2022qxg}. For example, a detailed analysis using data and MC simulations showed that high-multiplicity pp collisions selected at forward and backward rapidities, and requiring a hard process at midrapidity, results in the distribution of particles with multi-jet topologies, consequently affecting the search for medium-induced jet modification in small system~\cite{ALICE:2023plt}.

Different event classifiers are proposed to reduce the existing biases in selectors based only on the forward multiplicity. These include transverse spherocity ($S_{0}$)~\cite{Ortiz:2015ttf,ALICE:2019dfi,ALICE:2023bga} and the relative transverse activity classifier (\RT)~\cite{Martin:2016igp}, both measured at midrapidity. Spherocity is used to isolate pp collisions characterized by dijet topologies, which are dominated by hard partonic scatterings. Furthermore, it can select events with a large number of partonic interactions that yield an isotropic distribution of charged particles in the transverse plane~\cite{Ortiz:2017jho}. Regarding spherocity, pp collisions with spherocity values near zero are characterized by a dijet topology and are dominated by hard partonic scatterings. In contrast, events with spherocity values close to unity have an isotropic particle distribution in the transverse plane and are dominated by multiparton interactions. In Ref.~\cite{ALICE:2023bga}, particle spectra are analyzed as a function of spherocity and multiplicity, both measured at midrapidity and forward rapidity. Using the midrapidity multiplicity estimator together with a selection based on spherocity, it is possible to select events with a relatively large difference between the \meanpt of jetty and isotropic events as opposed to the case when the forward multiplicity estimator is used. However, a potential bias from jets fragmenting into many low-\pt particles emerges when the selection of high event activity at midrapidity is made. 

The \RT classifier selects pp collisions based on their UE activity in the region perpendicular to the direction of the leading charged particle, i.e.\ the one with the highest \pt, in the event~\cite{Martin:2016igp}. This approach probes the structure of the underlying event by separating events with exceptionally large or small transverse activity with respect to the event-averaged mean. This classification is applied in recent measurements of charged and identified particle production by ALICE~\cite{ALICE:2023yuk}, where the particle production is investigated as a function of the UE activity~\cite{ALICE:2019mmy,ALICE:2023yuk}. Phenomenological studies have found that transverse activity is strongly correlated with the average number of MPIs. In the region perpendicular to the leading particle, the spectral shapes of all particle species harden with increasing UE activity, which could be an indication of selecting multijet topologies. These effects could be a consequence of a selection bias originating from initial- and final-state radiations~\cite{Bencedi:2021tst}.  

Efforts have been made to develop a new event selector with reduced biases from local multiplicity fluctuations in the pseudorapidity region where multiplicity is measured. A necessary condition is to have a large sensitivity to quantities at the partonic level, such as MPI. For example, Ref.~\cite{Ortiz:2020rwg} proposes an event activity estimator with strong sensitivity to soft multiparton interactions and color reconnection effects using machine-learning-inspired techniques. In this study, the ratio of the yield of charged pions in pp collisions with a large number of MPIs to that in MB collisions shows a pronounced peak in the intermediate-$p_{\mathrm{T}}$ region ($2<p_{\rm T}<8\,\mathrm{GeV}/c$), which is attributed to CR. At larger \pt, such a ratio is consistent with unity. These effects have not been observed with the existing event activity estimators. In this context, the present paper explores a novel event classifier called flattenicity, which combines information from both the azimuthal and polar (pseudorapidity) angles~\cite{Ortiz:2022mfv}. The \pt spectra are studied in events selected as a function of flattenicity, with and without a multiplicity preselection in pp collisions at $\sqrt{s}=13$\,TeV. This measurement aims to provide further insights into the underlying processes behind collective phenomena, and it gives a better understanding of the partonic dynamics of the collisions. Moreover, it offers valuable information needed to improve the accuracy of event generators in describing the soft-QCD regime in small collision systems.

The paper is organized as follows. The ALICE experimental setup is described in Sec.~\ref{subsec:ExpSetup}, focusing on the detectors which are relevant to the presented measurements. Section~\ref{subsec:Flat} introduces charged-particle flattenicity and discusses its properties. Section~\ref{sec:EvtTrkSel} discusses the analyzed data samples, the details of the event and track selection criteria, the event classification, as well as the analysis techniques to measure the $p_{\rm T}$ spectra for the different particle species. Sections~\ref{sec:corrections} and~\ref{sec:systematic_uncertainties} outline the correction procedures and the estimation of systematic uncertainties. The results are presented and discussed in Sec.~\ref{sec:results}, including comparisons to Monte Carlo model predictions. Finally, Sec.~\ref{sec:Conclusions} gives the summary and draws the conclusions.

\section{Experimental apparatus}\label{subsec:ExpSetup}

A detailed description of the ALICE detector can be found in Ref.~\cite{ALICE:2008ngc}. The relevant detectors for the present analysis are the Inner Tracking System (ITS)~\cite{ALICE:2010tia}, the Time Projection Chamber (TPC)~\cite{Alme:2010ke}, the Time-Of-Flight (TOF) detector~\cite{Akindinov:2013tea}, and the V0 detectors~\cite{ALICE:2013axi}. These detectors are located in the central barrel surrounded by a solenoidal magnet, providing a homogeneous $B=0.5~\mathrm{T}$-magnetic field along $z$. 
The barrel includes a set of tracking detectors: the six-layer silicon \ITS detector surrounding the beam pipe, the large-volume ($5~\mathrm{m}$ length, $0.85~\mathrm{m}$ inner radius and $2.8~\mathrm{m}$ outer radius) cylindrical \TPC, and the \TOF detector. The \ITS is the innermost detector, covering the pseudorapidity region $|\eta|<0.9$. The two innermost layers are silicon pixel detectors (\SPD), followed by two intermediate layers composed of silicon drift detectors (\SDD), and finally, the two outermost layers are silicon strip detectors (\SSD). The \ITS measures the position of the primary collision vertex, the impact parameter of the tracks, and improves considerably the track-\pt resolution at high-\pt. The \TPC is the main detector for tracking and particle identification, covering the pseudorapidity range $|\eta|<0.8$ with full-azimuth coverage. With the measurement of drift time, the \TPC provides three-dimensional space-point information for each charged track, with up to 159 tracking points. Tracks originating from the primary vertex can be reconstructed down to $p_{\mathrm{T}}\sim\mevc{100}$~\cite{ALICE:2014sbx}, albeit with a lower tracking efficiency for identified charged hadrons below $\pt=\mevc{200}$. The \TPC provides charged-hadron identification via measurement of the specific energy loss \dEdx in the gas, with a resolution of ${\sim}5\%$ in pp collisions~\cite{ALICE:2014sbx}. The \TOF detector is a cylindrical array of multi-gap resistive plate chambers that surrounds the \TPC and covers the pseudorapidity range $|\eta|<0.9$ with full azimuthal acceptance. The time-of-flight is measured as the difference between the particle arrival time and the collision time of the event. The total time resolution, including the resolution on the collision time, is about $90~\mathrm{ps}$ in pp collisions. It enables particle identification up to about $\pt=\gevc{3}$~\cite{ALICE:2020nkc,ALICE:2023yuk}. In addition, the central barrel includes the V0 detectors. They are composed of two scintillator arrays placed along the beam axis ($z$) on each side of the nominal interaction point at $z =340~\mathrm{cm}$ and $z=-90~\mathrm{cm}$, covering the pseudorapidity regions $2.8<\eta<5.1$ (V0A) and $-3.7<\eta<-1.7$ (V0C), respectively. Each of the V0 arrays is segmented into four rings in the radial direction, and each ring is divided into eight sections in the azimuthal direction. This results in a lattice of $N{_{\mathrm{cell}}}=64$ cells. The amplitudes of V0A and V0C detector signals are proportional to charged-particle multiplicity, and their sum is denoted as V0M, used in event classification.  The V0 detectors provide the interaction trigger, and it is also used for beam-induced background suppression. Furthermore, it is employed for event classification based on multiplicity and flattenicity (see Sec.~\ref{subsec:Flat}). 

\section{Charged-particle flattenicity}\label{subsec:Flat}

Charged-particle flattenicity $(\rho)$ is measured on an event-by-event basis using the deposited energy registered in each cell of the \VZERO detector. The energy deposit in a given cell $i$ is proportional to the multiplicity of primary charged particles $(N_{\rm ch}^{{\rm cell},i})$.  Flattenicity is defined as follows~\cite{Ortiz:2022mfv}
 \begin{equation}
	\rho=\frac{\sqrt{\sum_{i=1}^{64}{(N_{\rm ch}^{{\rm cell},i} - \langle N_{\rm ch}^{\rm cell} \rangle)^{2}}/N{_{\mathrm{cell}}^{2}}}}{\langle N_{\rm ch}^{\rm cell} \rangle}\,,
\label{eq:flattenicity}
\end{equation}
where, $N_{\rm ch}^{{\rm cell},i}$ is the particle multiplicity in the $i$-th cell and $\langle N_{\rm ch}^{\rm cell} \rangle$ is the average over the total number of 64 cells per event. Flattenicity is therefore a measurement of the local multiplicity fluctuations in the V0 detector, small fluctuations are associated with $\rho \rightarrow 0$. It is demonstrated in Ref.~\cite{Ortiz:2022mfv} that $\rho$ is a robust observable against variations in the size of the cell. The values of $\rho$ range between 0 and 1. To associate flattenicity with other event shape observables, e.g. spherocity~\cite{ALICE:2019dfi, ALICE:2023bga}, the results are presented as a function of $1-\rho$.  Based on \py simulations, multijet topologies, that are produced by MPI, yield small flattenicity values ($1-\rho\rightarrow 1$), whereas pp collisions with a few MPIs have large flattenicity values ($1-\rho\rightarrow 0$). As a consequence, the lower bound of \orho aims at selecting ``soft'' pp collisions (including diffractive events), which, on average, produce a lower number of high-\pt hadrons compared with the inclusive (\orho–integrated) distribution, thereby making the \pt spectra softer. In contrast, the upper bound of \orho is associated with events with spherical topologies that contain particles from several multiparton interactions. By definition, flattenicity is a multiplicity-dependent quantity. Low-flattenicity events ($1-\rho\rightarrow 1$) have large event activity (i.e.\ large number of MPIs), and therefore rich QCD dynamics; this scenario can be reached in HM events~\cite{ALICE:2023bga}. On the contrary, the high-flattenicity limit ($1-\rho\rightarrow 0$) is associated with low-multiplicity events, mostly pp collisions with a few MPIs. These effects can be factorized by performing the event classification using a double-differential selection based on both multiplicity and flattenicity. 

\begin{figure*}[t]
\begin{center}
\includegraphics[width=0.66\textwidth]{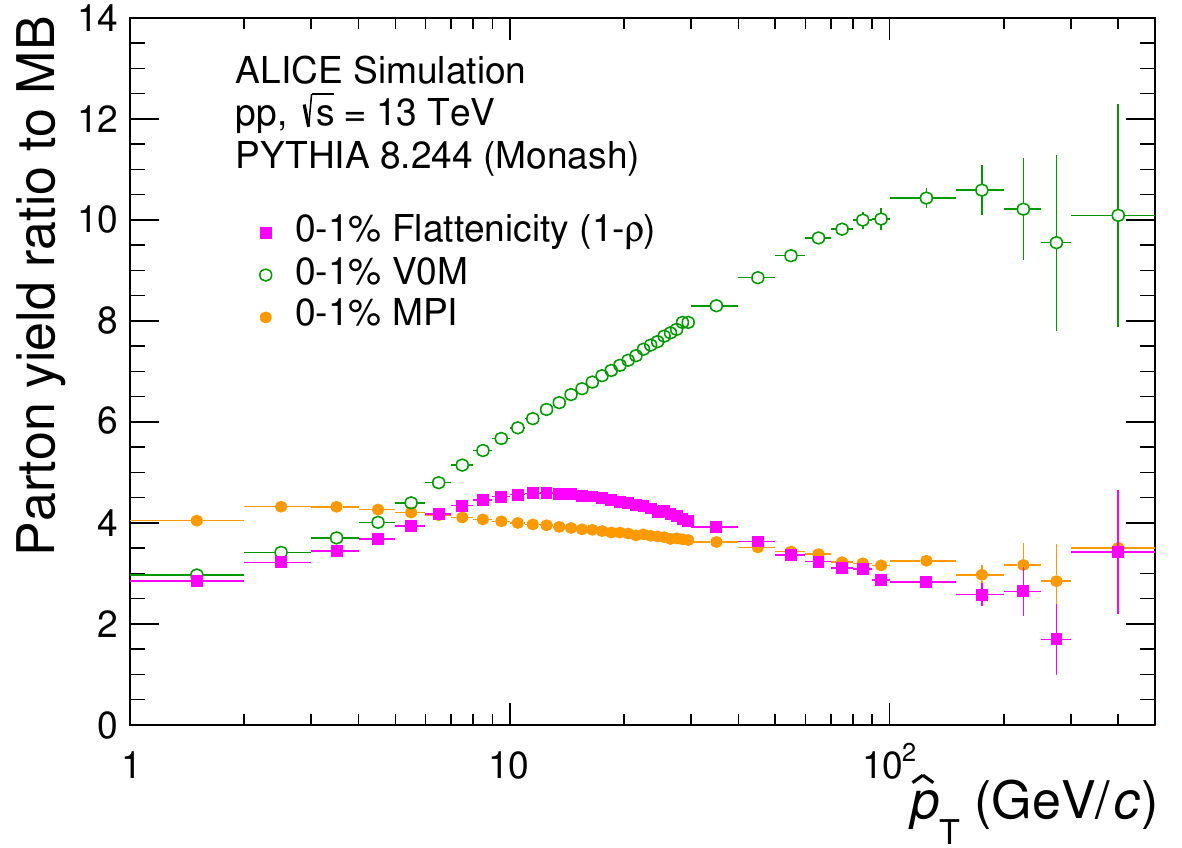}
\caption{Ratios of parton yields for the events with the 0--1\% highest activity according to various event activity measures to that without any event selection as a function of the parton transverse momentum (${\hat p}_{\rm T}$). The results are for pp collisions at \sppt{13} simulated with \py Monash 2013 tune.}
\label{fig:mc1}  
\end{center}
\end{figure*}

One can avoid trivial auto-correlation effects by measuring flattenicity in the forward rapidity region and the \pt spectra of charged particles in the midrapidity region. Studies based on \py showed that the calculation of flattenicity in the \VZERO acceptance (rather than at midrapidity) enhances the sensitivity to the global shape of the event~\cite{Ortiz:2022mfv}. As discussed earlier, there is a trivial correlation between MPI and the hardness of the collision. The larger the number of MPIs, i.e.\ collisions with small impact parameters, the larger the likelihood to produce a harder parton-parton scattering. However, selecting pp collisions based on their multiplicity biases the sample toward local multiplicity fluctuations originating from jets which yield a non-trivial effect. This is illustrated in Fig.~\ref{fig:mc1}, which shows the distribution of outgoing parton transverse momenta ($\hat{p}_{\rm T}$) in pp collisions simulated with \py with the highest 0-1\% event activity normalized to that in MB pp collisions. The event selection is done using the number of MPIs, flattenicity, and the V0M multiplicity estimator. The event selection based on MPI yields a ratio that is nearly flat up to ${\hat p}_{\rm T} \approx 10$\,GeV$/c$, and, from that  ${\hat p}_{\rm T}$ onward, followed by a slightly decreasing trend. For ${\hat p}_{\rm T}>30$\,GeV$/c$, a similar ratio as a function of ${\hat p}_{\rm T}$ is obtained when the event selection is performed using flattenicity. In contrast, the event selection based on V0M multiplicity yields a ratio that increases with  ${\hat p}_{\rm T}$. Overall, the flattenicity-dependent results are closer to the MPI-dependent results.

Given the sensitivity of the flattenicity to MPI, the selection of events with $1-\rho\rightarrow 1$ can enhance the color reconnection effects, which are more pronounced in collisions with a higher number of MPIs~\cite{OrtizVelasquez:2013ofg}. Color reconnection is expected to make a connection between the event activity in the forward region and the midrapidity region where the actual particle \pt spectra are measured~\cite{Kundu:2019ajc}. To test this assumption, a quantity $Q_{\mathrm{pp}}$ can be defined that demonstrates the evolution of the $p_{\mathrm{T}}$-spectral shapes with flattenicity
\begin{equation}
Q_{\mathrm{pp}} = \frac{1/\langle \mathrm{d}N_{\mathrm{ch}}/\mathrm{d}\eta \rangle_{\orho}}{1/\langle \mathrm{d}N_{\mathrm{ch}}/\mathrm{d}\eta \rangle_{\mathrm{MB}}} \frac{(\mathrm{d}^{2}N/ \mathrm{d}y\mathrm{d}\pt)_{\orho}}{(\mathrm{d}^{2}N/ \mathrm{d}y\mathrm{d}\pt)_{\mathrm{MB}}}~.
\end{equation}
The $Q_{\mathrm{pp}}$ quantity is given by the ratio of the particle yield measured in a given $1-\rho$ class to the yield measured in MB pp collisions. The $Q_{\mathrm{pp}}$ ratio is scaled by the ratio of average charged-particle pseudorapidity density measured in $|\eta| < 0.8$ for a given flattenicity event class to that for the MB event class $\langle \mathrm{d}N_{\mathrm{ch}}/\mathrm{d}\eta \rangle_{\orho}/\langle \mathrm{d}N_{\mathrm{ch}}/\mathrm{d}\eta \rangle_{\mathrm{MB}}$ that, according to \py, is proportional to the average number of MPIs. If a pp collision in a given flattenicity class behaved like a simple superposition of independent semi-hard parton-parton scatterings, the $Q_{\mathrm{pp}}$ would approach unity. 
\begin{figure}[t]
	\centering
	\hspace{0cm}
    \includegraphics[width=1\textwidth]{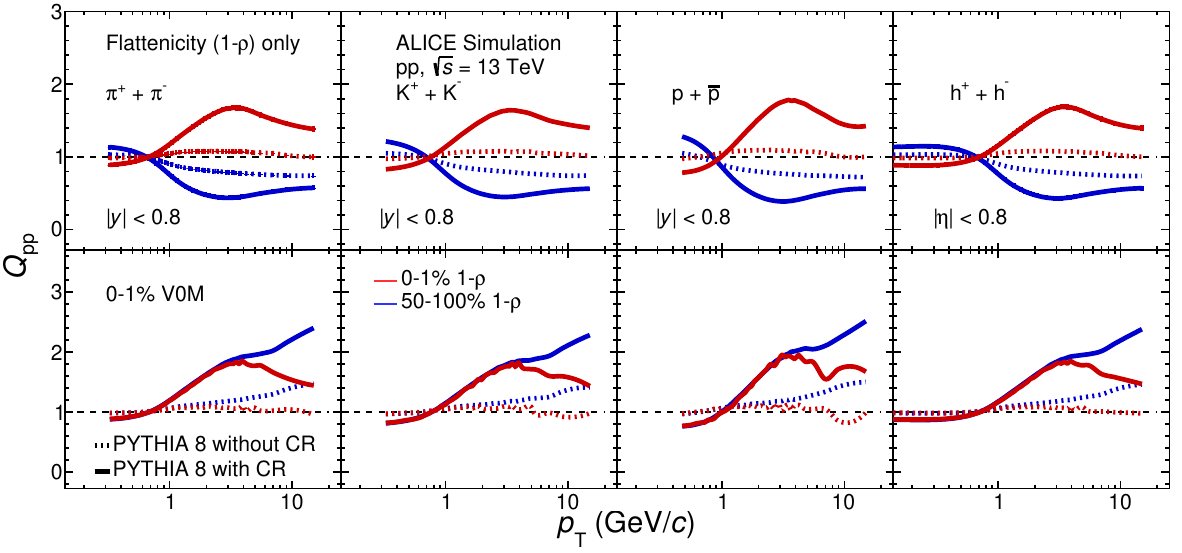}
	\hspace{0cm}
	\caption{The \Qpp ratio of \pion, \kaon, \pr, and \hadrons for the 0--1\% and 50--100\% flattenicity $(\orho)$ event classes (top row), and for HM events (0--1\% V0M) in the same \orho event classes (bottom row) simulated using \py with and without CR. There is a $|y|<0.8$ ($|\eta|<0.8$) cut in the rapidity (pseudorapidty) of identified (unidentified) particles. The shaded bands around the lines represent the statistical uncertainty.}
	\label{fig:rpp_only_models}
\end{figure}

Figure~\ref{fig:rpp_only_models} shows \Qpp of \pion, \kaon, \pr, and \hadrons for two extreme limits of flattenicity, the 0--1\% and 50--100\% \orho event classes, simulated with \py (with and without CR). There is a $|y|<0.8$ ($|\eta|<0.8$) cut in the rapidity (pseudorapidty) of identified (unidentified) particles. The \Qpp is around unity for 0--1\% \orho events without color reconnection regardless of particle species or event activity selection based on V0M~\cite{Ortiz:2020rwg}. This feature results from the sum of incoherent parton–parton collisions. Moreover, the \Qpp in the 50--100\% \orho class shows a slight decrease with increasing \pt because this type of pp collisions involve smaller momentum transfers than in MB pp collisions. On the other hand, the inclusion of the CR mechanism causes a deviation from unity: a ``bump'' structure and a dip emerge in the intermediate-\pt range ($1<p_{\mathrm{T}}<8~\mathrm{GeV}/c$) for the 0--1\% and 50--100\% \orho event classes, respectively. At higher \pt, the ratios approach unity like in the analogous ${\hat p}_{\rm T}$ plot shown in Fig.~\ref{fig:mc1}. The bottom row of Fig.~\ref{fig:rpp_only_models} shows the corresponding results for the HM event class. The results reveal similar features in the case of 0--1\% \orho. However, the $Q_{\mathrm{pp}}$ in the 50--100\% \orho event class increases over the entire $p_{\mathrm{T}}$ range. It is important to note that this effect was also seen for V0M-only event selections~\cite{ALICE:2023yuk,ALICE:2020nkc}, and it is a consequence of jet fragmentation bias~\cite{ALICE:2019dfi}. Despite the fact that the flattenicity is closely related to the event multiplicity ($\rho\propto 1/\sqrt{N_{\rm ch}}$), the observed features in the pp collisions with $\rho$ close to zero go beyond those obtained using a simple high-multiplicity selection. Moreover, the events with $\rho$ close to zero can be associated with collisions with many MPIs.


\section{Event and track selection}\label{sec:EvtTrkSel}

The present study uses a minimum-bias data sample from pp collisions at $\sqrt{s}=13$\,TeV collected between 2016 and 2018 during the Run 2 data-taking period of the LHC. The minimum-bias events are selected by the requirement of a charged-particle signal in both V0 detectors. Contamination from beam-induced background events is removed offline by using the timing information of the V0 detectors and by taking into account the correlation between the number of tracklets (short track segments reconstructed using only \SPD information) and the number of SPD clusters. Events are required to have a vertex position along the beam axis within $|z|<10~\mathrm{cm}$, where $z=0$ corresponds to the center of the detector. A selection criterion based on the offline reconstruction of multiple primary vertices in the \SPD is applied to remove contamination from pile-up events in the same bunch crossing~\cite{ALICE:2008ngc}. Furthermore, events with multiple interaction vertices reconstructed are rejected. After the offline rejection, the remaining pile-up has a negligible impact on the final results.

 
\begin{table}[t]
\caption{Average charged-particle multiplicity density \avdndeta in $|\eta|<0.8$ measured in different flattenicity event classes for 
multiplicity-integrated (V0M percentile 0--100\%) and high-multiplicity (V0M~percentile~0--1\%) events. The \avdndeta is measured by integrating the fully corrected \pt spectra of charged particles. The reported uncertainties correspond to the systematic contributions. Statistical errors are negligible compared to the systematic ones.}
\begin{center}
\begin{tabularx}{0.8\textwidth}{@{}lYYYY@{}}
\hline
\multicolumn{5}{c}{Multiplicity-integrated (V0M percentile 0--100\%)} \\
\hline
Class name & I & II & III & IV\\
\hline
$1-\rho $ percentile & 0--1\% & 1--5\% & 5--10\% & 10--20\% \\
$\langle \mathrm{d}N_{\mathrm{ch}}/\mathrm{d}\eta \rangle$ & $22.2\pm 0.7$ & $18.2\pm 0.5$ & $15.3\pm 0.5$ & $12.6\pm 0.4$ \\
\hline
\hline
Class name & V & VI & VII & VIII \\
\hline
$1-\rho $ percentile  & 20--30\% & 30--40\% & 40--50\% & 50--100\% \\
$\langle \mathrm{d}N_{\mathrm{ch}}/\mathrm{d}\eta \rangle$ & $10.0\pm 0.3$ & $8.06\pm 0.19$ & $6.47\pm 0.13$ & $3.51\pm 0.04$ \\
\hline
\hline \\[-8pt]
\multicolumn{5}{c}{V0M~percentile~0--1\%} \\
\hline
Class name & I & II & III & IV \\
\hline
$1-\rho $ percentile & 0--1\% & 1--5\% & 5--10\% & 10--20\% \\
$\langle \mathrm{d}N_{\mathrm{ch}}/\mathrm{d}\eta \rangle$ & $31.2\pm 0.5$ & $29.8\pm 0.4$ & $29.0\pm 0.4$ & $28.1\pm 0.4$ \\
\hline
\hline
Class name & V & VI & VII & VIII \\
\hline
$1-\rho $ percentile  & 20--30\% & 30--40\% & 40--50\% & 50--100\% \\
$\langle \mathrm{d}N_{\mathrm{ch}}/\mathrm{d}\eta \rangle$ & $27.4\pm 0.5$ & $26.7\pm 0.5$ & $26.1\pm 0.6$ & $24.0\pm 0.9$ \\
\hline
\hline
\end{tabularx}
\end{center}
\label{tab:flat_dNdEta}
\end{table}

\begin{figure*}[t]
\begin{center}
\includegraphics[width=0.7\textwidth]{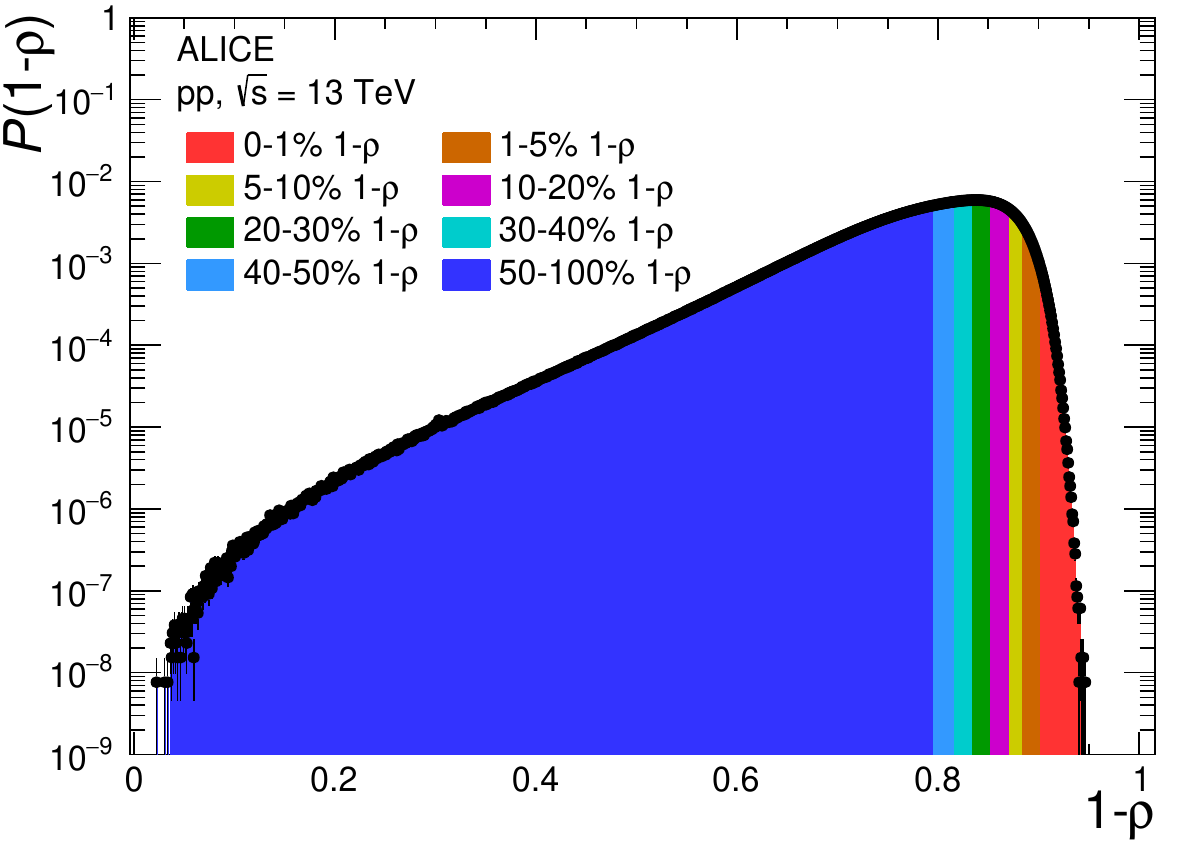}
\caption{Measured flattenicity probability distribution using the minimum bias sample. The colored areas represent the different percentile classes used in the measurement of the \pt spectra.}
\label{fig:flat_dist}  
\end{center}
\end{figure*}

The MB events are further required to have at least one charged particle produced in the pseudorapidity interval $|\eta|<1$. This class of events is referred to as INEL$>$0 and corresponds to about 75\% of the total inelastic cross section~\cite{ALICE:2012fjm,ALICE:2019avo,ALICE:2020nkc}. This study exploits about $1.64\times10^{9}$ selected minimum-bias pp collisions, corresponding to an integrated luminosity of about $21\,\rm{nb}^{-1}$. The multiplicity is classified in V0M percentiles, where 0--1\% corresponds to the highest (0--1\% V0M) multiplicity events. This HM event class will be used throughout the paper. The event selection based on flattenicity uses a procedure similar to that performed with the V0M multiplicity estimator. The simulated and the measured flattenicity distributions are divided into classes with different percentiles of the corresponding distribution. The measured flattenicity probability distribution with the minimum-bias sample and its division into percentiles is shown in Fig.~\ref{fig:flat_dist}. The percentiles used for the measurement of the \pt spectra of charged and identified particles and the corresponding average charged-particle pseudorapidity densities $\langle {\rm d}N_{\rm ch}/{\rm d}\eta \rangle$ measured within $|\eta| < 0.8$ are listed in Tab.~\ref{tab:flat_dNdEta}, where the values for 0--1\% V0M event class are also reported. Roman numerals represent the labeling convention for these percentiles, similar to what was reported in earlier ALICE publications~\cite{ALICE:2019dfi,ALICE:2023bga,ALICE:2023yuk}. The $\langle {\rm d}N_{\rm ch}/{\rm d}\eta \rangle$ is measured by integrating the fully corrected \pt spectra of charged particles. The details regarding the measurement of \avdndeta are given in Sec.~\ref{sec:systematic_uncertainties}. The flattenicity-integrated values of $\langle {\rm d}N_{\rm ch}/{\rm d}\eta \rangle$, i.e.\ the minimum-bias as well as the 0--1\% V0M values, are taken from Ref.~\cite{ALICE:2020swj}. A clear correlation between $\rho$ and $\langle {\rm d}N_{\rm ch}/{\rm d}\eta \rangle$ is observed: the 50--100\% \orho event class has lower $\langle {\rm d}N_{\rm ch}/{\rm d}\eta \rangle$ values than the 0--1\% \orho event class. 

The transverse momentum spectra are measured with primary charged particles~\cite{ALICE_PUBLIC_2017_005}. Charged particles are reconstructed using information from the \ITS and \TPC detectors within the pseudorapidity interval, $|\eta|<0.8$. The track selection criteria closely follow those used in Ref.~\cite{ALICE:2018vuu}. In particular, tracks are required to have clusters on at least 70 \TPC pad rows. They are also required to have at least two hits in the \ITS, out of which at least one is in the \SPD layers. The fit quality for the \ITS and \TPC track points must satisfy $\chi^{2}_{\mathrm{ITS}}/N_{\mathrm{hits}} < 36$ and $\chi^{2}_{\mathrm{TPC}}/N_{\mathrm{clusters}} < 4$, where $N_{\mathrm{hits}}$ and $N_{\mathrm{clusters}}$ are the number of hits in the \ITS and the number of clusters in the \TPC associated with the track, respectively. A 2-cm cut on the distance-of-closest approach (DCA) to the reconstructed primary vertex in the $z$-direction ($\mathrm{DCA}_{z}$) is applied to limit the contamination from secondary particles. Furthermore, a \pt-dependent selection on the $\mathrm{DCA}_{xy}$ in the plane perpendicular to the beam axis is applied.

\begin{table}[t]
\caption{The name of the analysis technique and the transverse momentum ranges in which \pion, \kaon, \pr are measured.}
\centering
  \begin{tabular}{c c c c}
  \toprule
  Analysis technique		             &  & {$p_{\rm T}~(\rm{GeV/}c)$} ranges &    \\ [1pt]
  \midrule \\[-15pt]
                                       & $\pi$ & $\mathrm{K}$ & $\mathrm{p}$       \\ [1pt]
  {TPC}               	             & $0.3-0.7$ & $0.3-0.6$ 	& $0.45-1$	     \\ [2pt]
  {TOF}               	             & $0.7-3$   & $0.6-3$      & $1-3$	         \\ [2pt]
  {TPC rel. rise}                      & $2-20$    & $3-20$ 	    & $3-20$	   	 \\ [2pt]
  \bottomrule
  \end{tabular}
\label{tab:pt_ranges}
\end{table}

The particle identification (PID) is performed using the standard techniques explained in previous ALICE publications~\cite{ALICE:2020nkc}. Table~\ref{tab:pt_ranges} lists the names of the analysis techniques and the \pt range in which the spectra are measured. Below $\pt=\gevc{1}$, PID is performed on a track-by-track basis~\cite{ALICE:2020nkc}. Up to $\pt=\gevc{3}$, the yield of \pion, \kaon, and \pr is extracted from the information provided by the \TOF detector~\cite{ALICE:2023yuk}. Finally, the TPC relativistic rise method is employed in the region $2<p_{\rm T}<20$\,GeV/$c$, where the yield is measured by fitting the ${\rm d}E/{\rm d}x$ spectrum in the relativistic rise regime of the Bethe--Bloch curve as described in Ref.~\cite{ALICE:2014juv}.

\section{Corrections}\label{sec:corrections}

The fully corrected \pt spectra are obtained using standard methods~\cite{ALICE:2019hno}. The set of corrections includes the limited acceptance and tracking inefficiency, \TPC-\TOF matching inefficiency (only where the \TOF measurement is used for PID), secondary particle contamination, and event and signal losses. All corrections are calculated using events simulated with PYTHIA~8.2 tune Monash 2013, hereafter referred to as \py~\cite{Skands:2014pea}. The simulated particles are subsequently propagated through a simulation of the ALICE detector using the \gea transport code~\cite{Geant3}. The simulated particles are reconstructed using the same algorithms as for the data. Since the tracking and matching efficiencies, and the contamination of secondary particles show little or no dependence on the event multiplicity, the minimum-bias result is used for all the multiplicity and flattenicity classes. The tracking inefficiency of unidentified charged particles takes into account the measured particle composition of the charged spectrum as described in ~\cite{ALICE:2020jsh}. The residual contamination from secondary particles (the products of weak decays and particles produced due to interactions with the detector material) is estimated by fitting the data $\mathrm{DCA}_{xy}$ distributions in \pt bins using Monte Carlo templates describing the contribution of primary and secondary particles~\cite{ALICE:2019hno,ALICE:2023yuk}. This correction amounts to 1\%, 10\%, and 3\% at $\pt \approx 0.5~\gevc$ for \pipm, \pr(\pbar), and charged hadrons, respectively. Finally, the spectra are corrected for event and signal losses, which take into account the trigger selection inefficiency~\cite{ALICE:2019avo}. Both corrections are relevant for low-multiplicity events. In particular, the signal-loss correction is the largest for events in the 50--100\% \orho class, it amounts up to $6\%$ at $\pt=0.3~\mathrm{GeV}/c$ and decreases to $1\%$ at $\pt=10~\mathrm{GeV}/c$. For the same class of events, the magnitude of the event loss correction is about $12\%$. The correction procedure is tested by performing a Monte Carlo closure test, which is defined as, $(\mathrm{d}^{2}N/\mathrm{d}\pt\mathrm{d}y)_{\orho,\mathrm{meas}}/(\mathrm{d}^{2}N/\mathrm{d}\pt\mathrm{d}y)_{\orho,\mathrm{gen}}$, where the numerator is the fully corrected \pt spectrum for a flattenicity class selected using the measured \orho per event, and the denominator is the \pt spectrum at generator level (no detector effects included) for the same flattenicity class using the true \orho. 
 

\section{Systematic uncertainties}\label{sec:systematic_uncertainties}

The total systematic uncertainty on the \pt spectra is estimated using standard procedures described in Refs.~\citep{ALICE:2023yuk,ALICE:2019dfi,ALICE:2020nkc}. The different sources of uncertainty are grouped into two disjoint classes: common uncertainties between the charged and identified-particle analyses, and analysis-specific uncertainties. The former class includes the uncertainties due to the vertex and track selections, event and signal loss corrections, \ITS-\TPC and \TPC-\TOF matching efficiencies, and Monte Carlo non closure. The systematic uncertainty on the \ITS-\TPC and \TPC-\TOF matching efficiencies is taken from Ref.~\cite{ALICE:2020nkc}. The quantification of the systematic uncertainty specific to the extraction of the identified-particle yield, and to the estimation of the secondary particle contamination correction is described in detail in Ref.~\citep{ALICE:2019hno}. The individual sources of uncertainty are summed in quadrature to obtain the total systematic uncertainty on the \pt spectra. Tables~\ref{tab:SystChrg} and~\ref{tab:syspikp} summarize the different sources of uncertainty in the charged and identified particle analyses. Below, only a brief description of the sources of systematic uncertainty, which depend on the flattenicity selection is given.

\begin{itemize}
    \item Monte Carlo (MC) non closure: This is measured as a function of the multiplicity and flattenicity selections. It is estimated by comparing the fully corrected \pt spectra with the spectra obtained in the MC simulation at the generator level. For the 50--100\% \orho class, the non closure has its maximum value of about $16\%$ at $\pt=\gevc{3}$, whereas, for the 0--1\% \orho class, its value is estimated to be 10\% at $\pt=\gevc{0.15}$, at which the value is largest. For the HM events (0--1\% V0M), the non closure is between about 3.5\% and 21\% for the 0--1\% \orho class, whereas it amounts to 12--19\% in the 50--100\% \orho class, depending on \pt. The main source of the uncertainty of the MC non closure is related to the effect of secondary particles that enter the measured flattenicity, which are not considered in the calculation of flattenicity in MC at the generator level.

    \item Event and signal loss corrections: These corrections have a modest dependence on the Monte Carlo event generator. Therefore, the \ep model is used to quantify a second set of corrections. These corrections depend on the multiplicity and/or flattenicity class, the transverse momentum, and the particle species. The difference between the corrections obtained with \py and \ep is assigned as the systematic uncertainty. In particular, the signal loss correction uncertainty for unidentified charged particles in the 50-100\% flattenicity class is between 0.6\% and 2.6\% over the entire \pt range and becomes negligible for the 0-1\% class. The event loss correction uncertainty, which depends only on the multiplicity and/or flattenicity class is about 0.6\% for the 50--100\% flattenicity class and negligible for the 0--1\% class.

\end{itemize}

The charged-particle pseudorapidity densities $\langle \mathrm{d}N_{\mathrm{ch}}/\mathrm{d}\eta \rangle$ (c.f.\ discussion in Sec.~\ref{sec:EvtTrkSel}), the \pt-integrated particle yields (\dndy), and the average transverse momenta (\meanpt) were calculated using the measured \pt distributions and their extrapolations based on L\'evy--Tsallis fits to unmeasured \pt regions, similar to what was done in previous measurements~\cite{ALICE:2013rdo,ALICE:2018pal,ALICE:2020nkc}. The fractions of extrapolated yields in the 0--1\% \orho class amount to $34\%, 14\%$, and $15\%$ for \pion, \kaon, and \pr, respectively. The variation of fit ranges and other fit functions (Boltzmann--Gibbs blast wave, $m_{\rm T}$-exponential, Fermi--Dirac, and Bose--Einstein) were considered to estimate the systematic uncertainties related to the procedure. The resulting variations in the $\langle \mathrm{d}N_{\mathrm{ch}}/\mathrm{d}\eta \rangle$, \dndy, and \meanpt values are incorporated into the systematic uncertainties. The total systematic uncertainties, for example for the 0--1\% \orho class, on the $\mathrm{d}N/\mathrm{d}y$ and \meanpt amount to 4.4\% and 3\% for \pion, 4\% and 2\% for \kaon, and 3.2\% and 2\% for \pr, respectively.

\begin{table*}[t]
\caption[]{Main sources and values of the relative systematic uncertainties on the \pt spectra of charged particles. They are given for three different \pt values. The abbreviation ``negl.'' indicates a negligible value.}

\centering
\label{tbl:systchrg}
\begin{tabular}{l c c c c c c}
\toprule
{$p_{\rm T}~(\rm{GeV/}c)$} 		& 0.15 &  & 3.0 &  & 10\textbf{}  & \\ [1pt]
  \hline \\[-8pt]
  \textbf{Source of uncertainty} \\ [1pt]
  {Vertex selection}        	 & 0.1\% 	&	& 0.1\% 	&	& 0.7\%	& 		\\ [2pt]
  {Track selection}         	 & 1.1\% 	&	& 0.5\% 	&	& 0.9\%	& 		\\ [2pt]
  {\ITS-\TPC matching efficiency}     	 & 2.0\% 	&	& 4.0\% 	&	& 5.0\%	& 		\\ [2pt]
  {Secondary particles}     	 & 1.1\% 	&	& negl. 	&	& negl.	& 		\\ [2pt]
  {$p_{\rm T}$ resolution} 	 & negl. 	&	& negl. 	&	& 0.1\%	& 		\\ [2pt]
  {Particle composition}    	 & 0.2\% 	&	& 1.5\% 	&	& 0.3\%	& 		\\ [2pt]
  {MC non-closure}    	     & 1.5\% 	&	& 15.9\% 	&	& 4.8\%	& 		\\ [2pt]
  \hline \\[-10pt]
  {Total}                   	 & 2.9\% 	&	& 16.5\% 	&	& 7.0\%	& 		\\ [2pt]
\bottomrule
\end{tabular}
\label{tab:SystChrg}
\end{table*}

\begin{table*}[!ht]\label{syst}
\caption{Summary of systematic uncertainties on the \pt spectra of \pion,  \kaon, and \pr. The uncertainties are shown for three different representative \pt values. The last two rows show the total systematic uncertainty on the \pt spectra and the \pt-differential particle ratios. The values of MC non-closure are given for the 0--1\% \orho class. }
\begin{tabularx}{\textwidth}{p{5.3cm}*{2}{Y}*{1}{Y | }*{2}{Y}*{1}{Y | }*{3}{Y}}
\toprule
    \textbf{Source of uncertainty} \\
    {\bf Common}  & \multicolumn{3}{c}{\pion} & \multicolumn{3}{c}{\kaon} & \multicolumn{3}{c}{\pr} \\
\hline
    \pt\ (\GeVc)   &   0.3 & 3 & 10   &   0.3 & 3 & 10 &   0.45 & 3 & 10 \\
    \hline

    \ITS--\TPC matching efficiency  & 1.4\% & 2.6\% & 5\% & 1.4\% & 2.6\% & 5\% & 1.4\% & 2.6\% & 5\%\\

    Vertex selection & 0.1\% & 0.1\% & 0.7\% & 0.1\% & 0.1\% & 0.7\% & 0.1\% & 0.1\% & 0.7\% \\

    Track selection & 0.7\% & 0.5\% & 1\% & 0.7\% & 0.5\% & 1\% & 0.7\% & 0.5\% & 1\% \\

    MC non closure & 10\%  & 9.3\% & 1.8\% & 10\% & 9.3\% & 1.8\% & 10\% & 9.3\% & 1.8\% \\

\hline
    {\bf Analysis-specific} & \multicolumn{3}{c}{\pion} & \multicolumn{3}{c}{\kaon} & \multicolumn{3}{c}{\pr} \\
\hline

    {\bf TPC}, \pt\ (\GeVc) & 0.3 & & 0.7 & 0.3 & & 0.6 & 0.45 & & 1\\
    \hline
    PID & 0.1\% & & 3\% & 1.5\% & & 7.8\% & 2.5\% & & 3.2\% \\
    Feed-down & 0.8\% & & 0.1\% & - & & - & 10\% & & 1\% \\

\hline\hline

    {\bf TOF}, \pt\ (\GeVc)  & 1 &  & 2   &   1 &  & 2  &   1 &  & 2    \\
    \hline
    PID & negl. & & 2.3\% & 1.4\% & & 7.2\% & negl. & & 0.9\% \\
    Feed-down & negl. & & negl. & - & & - & 1\% & & 0.1\% \\
    TOF matching efficiency & 3\% & & 3\% & 6\% & & 6\% & 4\% & & 4\% \\

\hline\hline

    {\bf TPC rel. rise}, \pt\ (\GeVc) & 3 & & 10 & 3 & & 10  &   3 & & 10    \\
    \hline
    PID & 1\% &  & 1.5\% & 10.3\% &  & 3.5\%  & 11.6\% & & 5.8\% \\
    Feed-Down & negl. & & negl. & - &  & - & 0.1\% & & 0.1\% \\

\hline\hline

   {\bf Total}  & \multicolumn{3}{c}{\pion} & \multicolumn{3}{c}{\kaon} & \multicolumn{3}{c}{\pr} \\
\hline
    \pt\ (\GeVc)   &   0.3 & 2 & 10   &   0.3 & 2 & 10 &   0.45 & 2 & 10 \\
    \hline
    Total & 10.1\% & 10.7\% & 5.5\% & 10.2\% & 15.4\% & 6.4\% & 14.4\% & 10.6\% & 8\% \\
\hline\hline
    {\bf Particle ratios}  & \multicolumn{3}{c}{} & \multicolumn{3}{c}{\ktopi} & \multicolumn{3}{c}{\ptopi} \\
\hline
    \pt\ (\GeVc)   &    &  &    &   0.3 & 2 & 10 &   0.45 & 2 & 10 \\
\hline
    Total & & & & 7\% & 4\% & 4.4\% & 10.4\% & 3.5\% & 4.7\% \\
    \bottomrule
  \end{tabularx}
  ~\newline
  \label{tab:syspikp}
\end{table*}

\section{Results and discussion}\label{sec:results}

This section describes the transverse momentum spectra, \dndy, \meanpt, \pt-differential particle ratios, and \pt-integrated particle ratios as a function of flattenicity and double-differentially as a function of flattenicity in HM (0--1\% V0M class) events. 

The results presented below are compared with theoretical predictions from QCD-inspired MC models. Besides the \py model introduced earlier, \ep~\cite{Werner:2005jf,Pierog:2013ria} is also used for comparisons. \ep is a two-component core-corona model: the high energy-density ``core'' region undergoes a collective expansion and hadronization including radial and longitudinal flow effects, whereas the low-density ``corona'' region is described by independent string fragmentation and hadronization.

\begin{figure}[t]
    \centering
    \hspace{0cm}
    \includegraphics[width=1\textwidth]{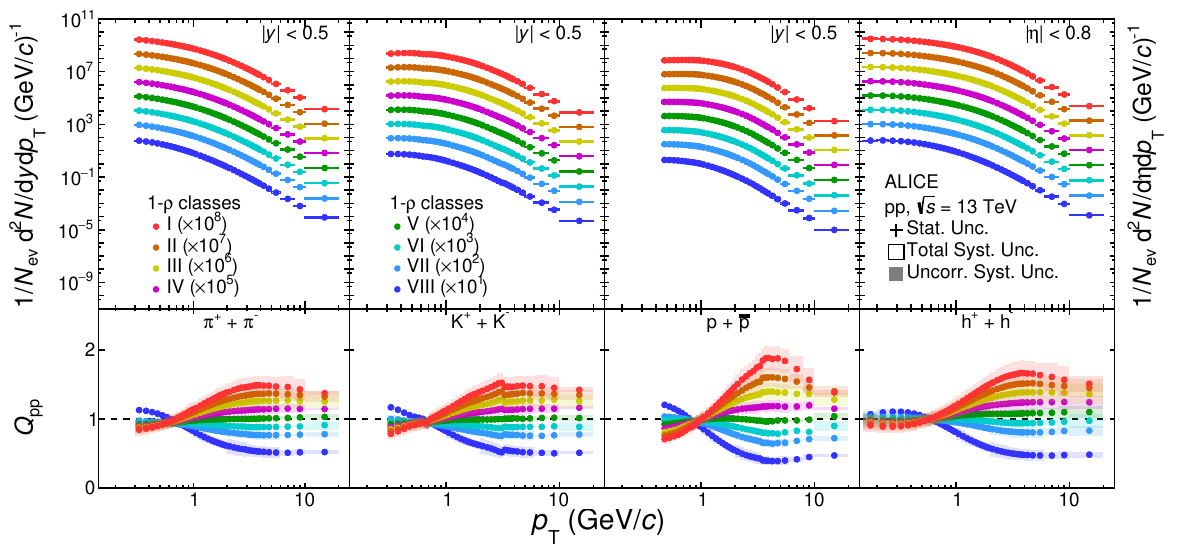}
    \hspace{0cm}
    \includegraphics[width=1\textwidth]{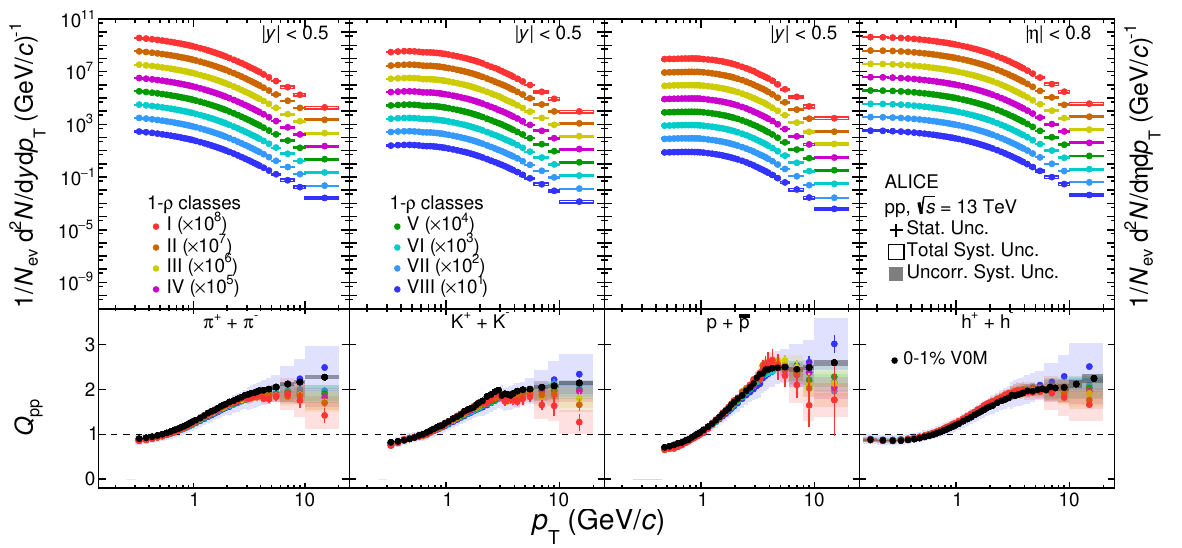}
    \hspace{0cm}
    \caption{Transverse momentum ($p_{\rm T}$) spectra of $\pi^{\pm}$, $\mathrm{K}^{\pm}$, $(\overline{\mathrm{p}})\mathrm{p}$, and $\rm{h}^{\pm}$ for different flattenicity event classes (top panels), and for HM events (0--1\% V0M) in the same flattenicity event classes (bottom panels). The spectra are scaled by powers of ten for better visibility. The yield of identified and unidentified particles is reported as a function of rapidity and pseudorapidity, respectively. The bottom panels in each figure show the $Q_{\rm pp}$ for the corresponding event classes. The statistical, total, and uncorrelated systematic uncertainties are represented with bars, boxes, and shaded areas around the data points, respectively. The flattenicity-integrated $Q_{\rm pp}$ values are taken from Refs.~\cite{ALICE:2019dfi,ALICE:2020nkc}.}
    \label{fig:spectra_with_rpp}
\end{figure}

The top part of Fig.~\ref{fig:spectra_with_rpp} shows the \pt spectra of \pion, \kaon, \pr, and \hadrons  as a function of charged-particle flattenicity. As discussed in Sec.~\ref{subsec:Flat}, the \Qpp ratio can be used to illustrate the sensitivity to MPI and CR effects. The \pt spectra of charged particles are used to derive the average charged-particle pseudorapidity densities $\avdndeta$. These values are reported in Tab.~\ref{tab:flat_dNdEta} in Sec.~\ref{sec:EvtTrkSel} and illustrate the implicit multiplicity dependence of flattenicity. The bottom panels of Fig.~\ref{fig:spectra_with_rpp} show the \pt dependence of $Q_{\mathrm{pp}}$ for the corresponding flattenicity classes. A clear development of a peak structure for the flattenicity event class I is observed for $1<p_{\mathrm{T}}<8~\mathrm{GeV}/c$. In contrast to previous measurements as a function of V0M multiplicity~\cite{ALICE:2019dfi,ALICE:2020nkc}, where similar ratios to \Qpp show an increasing trend with \pt for HM events, the \Qpp shows a hint for a gradual decrease at higher \pt for all flattenicity classes. This is consistent with the MC results, which suggest that flattenicity can be a potential observable to select HM pp collisions while minimizing the bias due to local multiplicity fluctuations. The bottom part of Fig.~\ref{fig:spectra_with_rpp} reports the results from a double-differential analysis, where HM events (0--1\% V0M) are first selected, and then a flattenicity classification is applied. The HM event class has, on average, three to four times larger $\langle{\rm d}N_{\rm ch}/{\rm d}\eta\rangle$ with respect to MB events. However, the $Q_{\mathrm{pp}}$ in the event class VIII increases over the entire $p_{\mathrm{T}}$ range. In addition, the \Qpp measurements from this double-differential study are compared with those obtained for V0M-only event selections~\cite{ALICE:2023yuk,ALICE:2020nkc}, shown as black markers. The \Qpp that depends only on the multiplicity selection is closer to that in the 50--100\% \orho class for the same multiplicity class. The increasing trend of the \Qpp can be attributed to a multiplicity-based selection biasing the sample toward collisions featuring fragmentation of hard partons.

\begin{figure}[t]
    \centering
    \hspace{0cm}
    \includegraphics[width=1\textwidth]{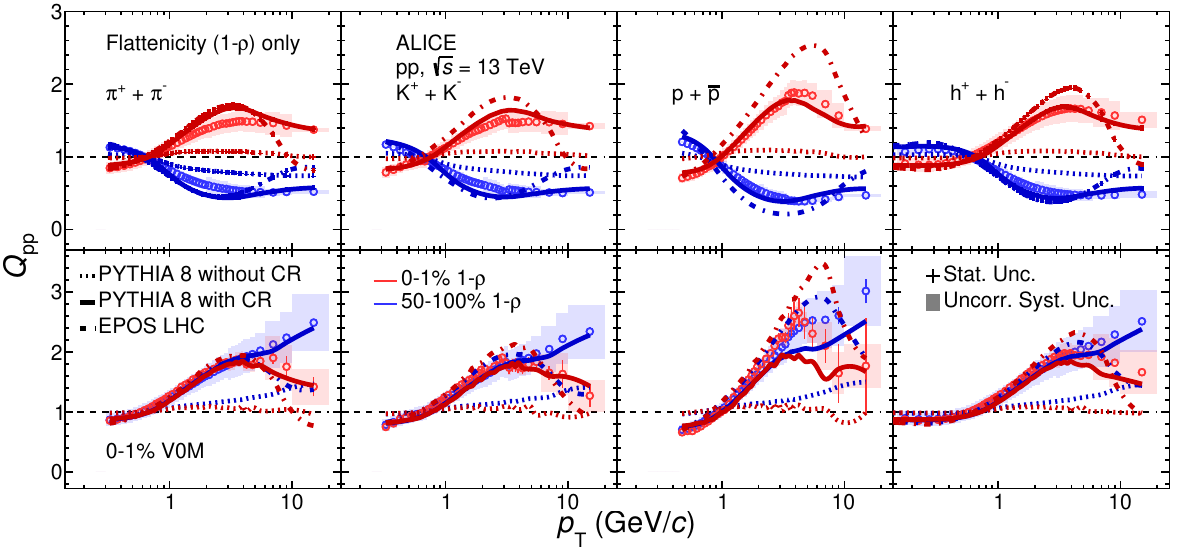}
    \hspace{0cm}
    \caption{The \Qpp ratio of \pion, \kaon, \pr, and \hadrons for the 0--1\% and 50--100\% \orho classes (top row) and for the same \orho classes in HM (0--1\% V0M class) events (bottom row). The data are compared with \py and \ep model predictions. The statistical and systematic uncertainties are represented with bars and shaded areas.}
    \label{fig:rpp_with_models}
\end{figure}

Figure~\ref{fig:rpp_with_models} shows the measured \Qpp ratios of \pion, \kaon, \pr, and \hadrons , and model predictions from \py~\cite{Skands:2014pea} (with and without CR) and \ep~\cite{Pierog:2013ria}. Here, only the extreme flattenicity selections are examined, 0--1\% and 50--100\% \orho. Results in the top row were obtained for multiplicity-integrated (0--100\% V0M) events, whereas those shown in the bottom row were produced for high-multiplicity (0--1\% V0M) events. The measured \Qpp ratios for low- and high-flattenicity selections intersect unity at $\pt\approx\gevc{0.5}$ regardless of particle species and multiplicity selection. The data deviates from unity, and it depends on both the flattenicity selection and \pt. The prediction based on \py without color reconnection effects (c.f.\ discussion in Sec.~\ref{subsec:Flat}) yields \Qpp ratios consistent with unity, and it is far from describing the data. On the contrary, the \py model with the Monash 2013 tune, that includes MPIs and CR effects, generally describes better the measurements of \pion, \kaon, \pr, and \hadrons in flattenicity event classes. The \ep model with parametrized collective hydrodynamics describes the data only partially (low-to-mid $p_{\rm T}$), while at high $p_{\rm T}$ it underestimates \Qpp for pions, kaons, and unidentified hadrons. In the double-differential analysis, \py with CR describes the data well for both flattenicity event classes, although it gives only a qualitative description for protons. 

\begin{figure}[t]
    \centering
    \hspace{0cm}
    \includegraphics[width=1\textwidth]{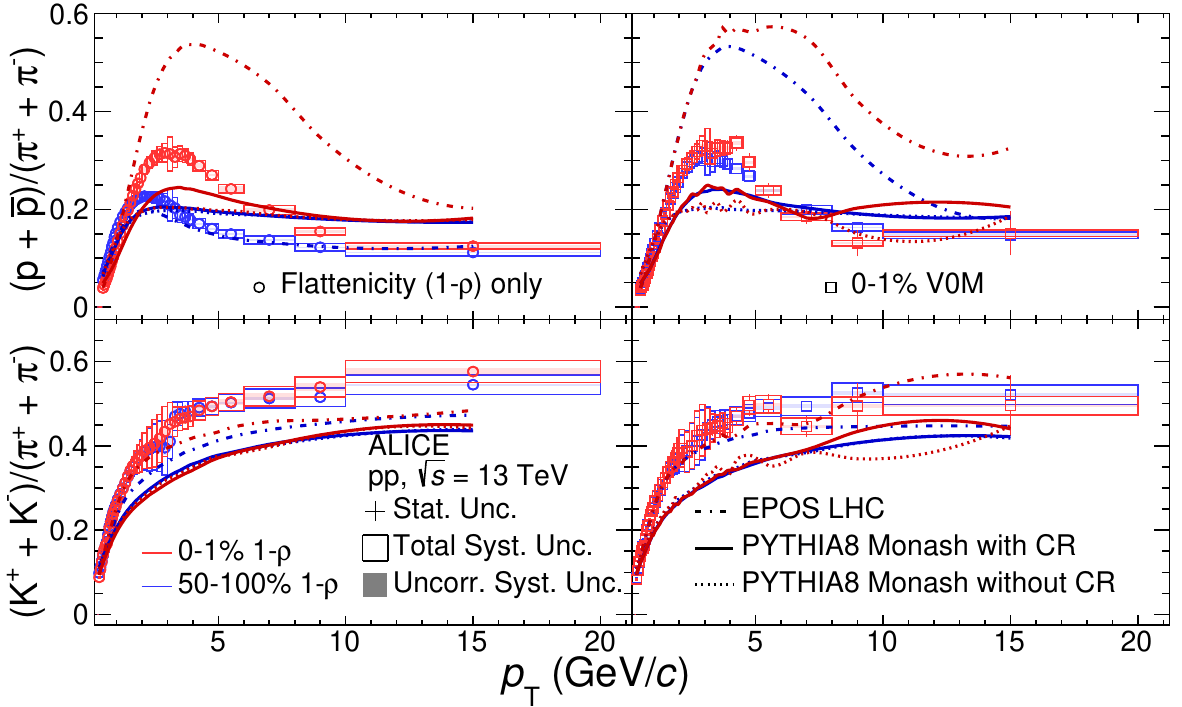}
    \hspace{0cm}
    \caption{The \pt-differential proton-to-pion (top row) and kaon-to-pion (bottom row) particle ratios for two extremes of flattenicity event classes, as indicated in the legends, are shown. Left and right columns include results for multiplicity-integrated and 0--1\% V0M event classes. The statistical, total, and uncorrelated systematic uncertainties are represented with bars, boxes, and shaded areas around the data points, respectively. The shaded regions around the model line represent the statistical uncertainties.}
    \label{fig:particle_ratios_with_models}
\end{figure}

Figure~\ref{fig:particle_ratios_with_models} shows the \pt-differential proton-to-pion (\ptopi), and kaon-to-pion (\ktopi) ratios for the two extremes of flattenicity: 0--1\% and 50--100\% \orho. Left and right columns include results for multiplicity-integrated and 0--1\% V0M event classes. The \ktopi ratio does not depend on flattenicity, neither in the multiplicity-integrated case nor in
high-multiplicity events. The same is true for the model calculations with \py. A higher \ptopi ratio is observed in the 0--1\% \orho class with respect to the 50--100\% \orho one for $2 \lesssim \pt \leq 10~\mathrm{GeV}/c$ when only a selection based on flattenicity is applied. This effect has already been reported in previous ALICE publications, where the particle production was measured as a function of event multiplicity~\cite{ALICE:2018pal,ALICE:2020nkc,Ortiz:2022mfv}. The \py model with color reconnection also predicts an enhanced baryon-to-meson ratio for the 0--1\% \orho class with respect to the 50--100\% \orho one. \ep predicts different \ptopi ratios as a function of flattenicity, and while it shows a very good agreement with the 50--100\% \orho class, it overestimates the data in the 0--1\% \orho one. For $\pt \geq 10~\mathrm{GeV}/c$, the measured \ptopi ratio between the two flattenicity classes is the same. However, the maximum in the highest \orho interval is shifted to the right with respect to the lowest \orho interval; this might be attributed to the jet hardening effect with increasing multiplicity. Finally, for the 0-1\% V0M multiplicity class, the \ptopi ratios do not exhibit a strong flattenicity dependence. This feature is replicated by \py with and without color reconnection effects, while \ep predicts trends that are not observed in the data. It is worth mentioning that a complementary analysis based on multiplicity and spherocity measured at midrapidity exhibits a strong event-shape dependence. A reduction of the particle ratios for jet-like events is observed~\cite{Ortiz:2017jho,ALICE:2023bga}. 

\begin{figure}[t]
    \centering
    \hspace{0cm}
    \includegraphics[width=0.48\textwidth]{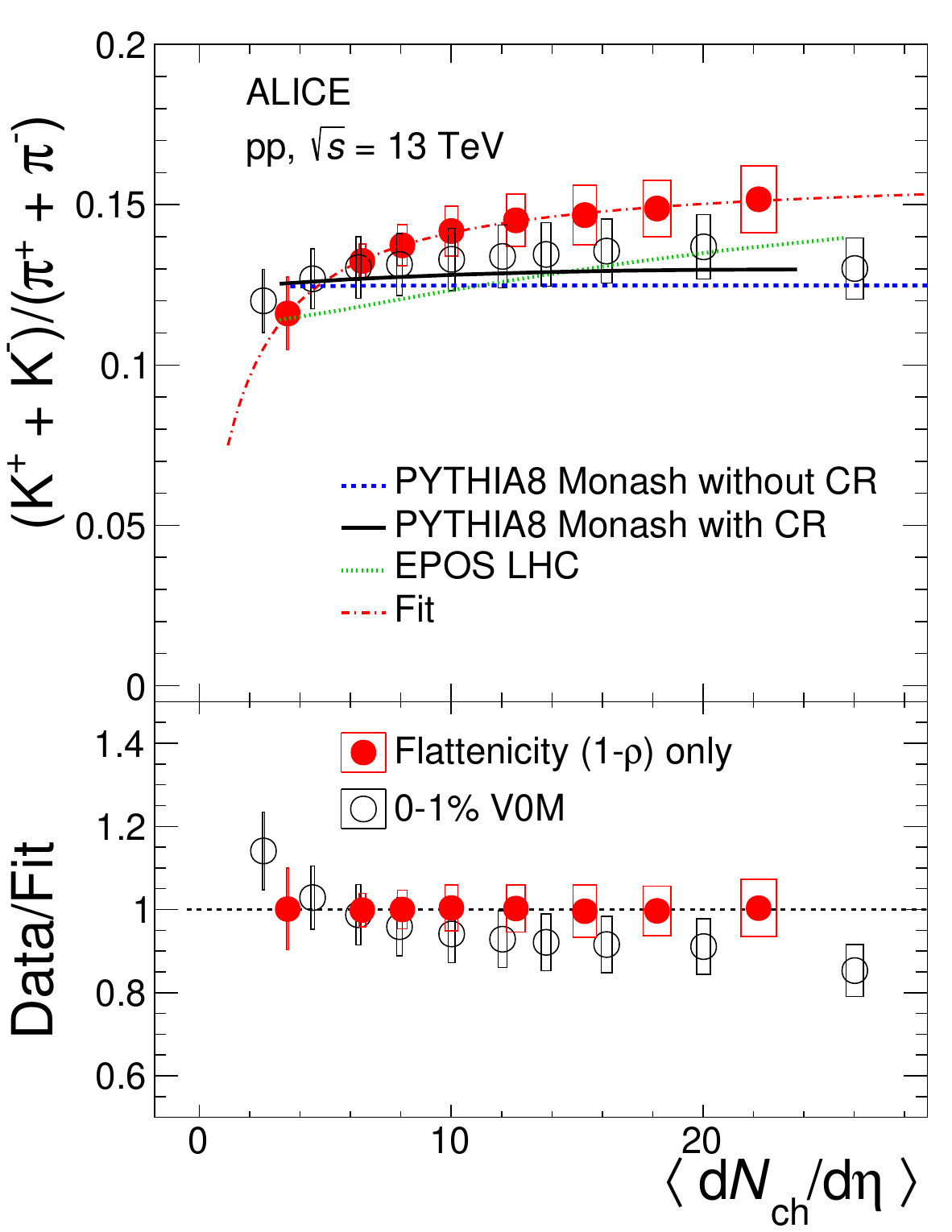}
    \hspace{0cm}
    \includegraphics[width=0.48\textwidth]{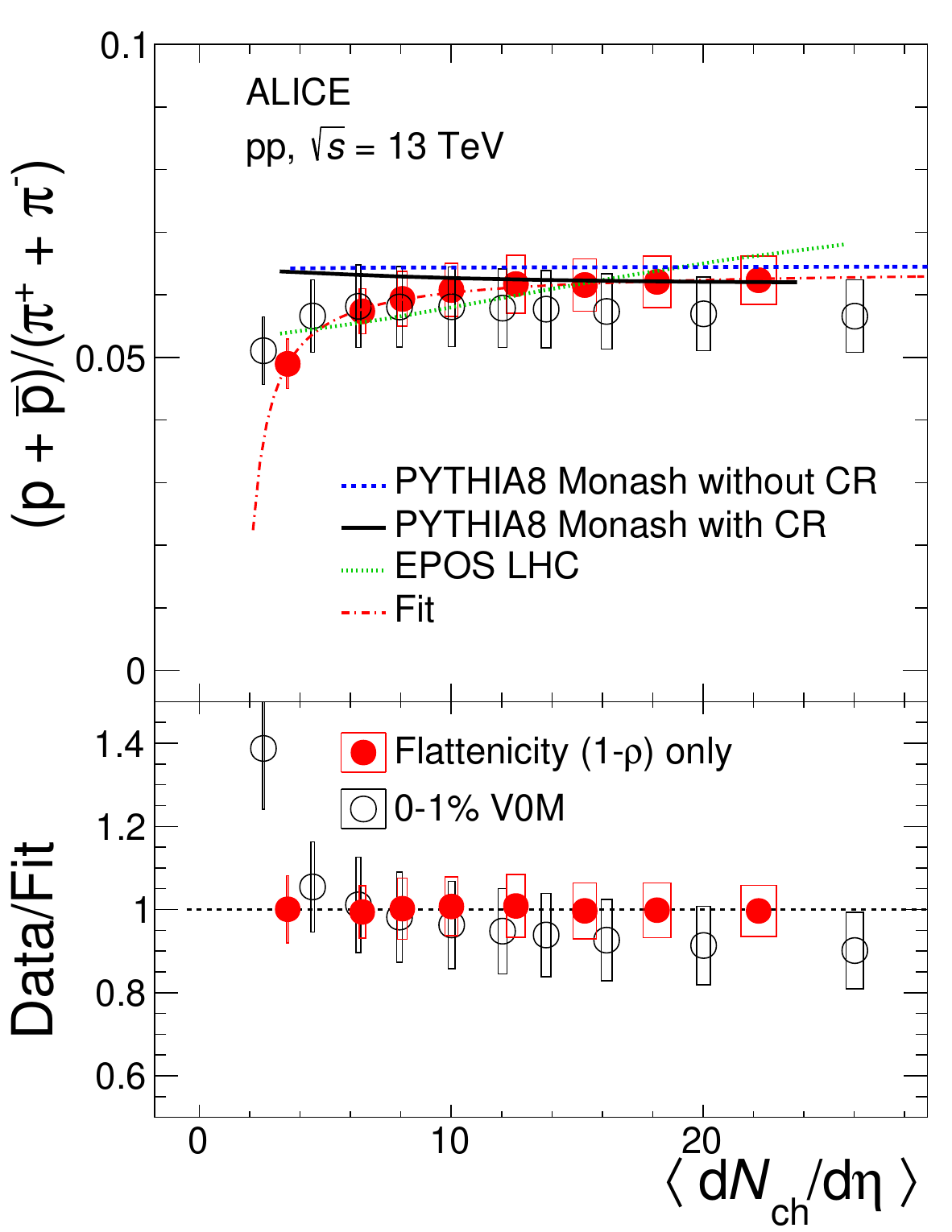}
    \caption{Top panels: Transverse momentum-integrated particle ratios as a function of the average charged-particle density, and predictions using the \py and \ep models. The model calculations and fits (shown with the red dashed lines) correspond to the flattenicity-dependent measurements. The fit uses the $a-b/(c-x)$ parameterization, where $a,b$, and $c$ are free fit parameters. Bottom panels: Data-to-fit ratios for both the flattenicity- and multiplicity-dependent measurements. The multiplicity-dependent results are taken from Ref.~\cite{ALICE:2020nkc}. The statistical and systematic uncertainties are shown with lines and empty boxes.}
    \label{fig:integrated_particle_ratios}
\end{figure}

\Cref{fig:integrated_particle_ratios} shows the \pt-integrated \ktopi and \ptopi ratios as a function of \avdndeta with a flattenicity-based selection only.  The measurements are compared with their counterparts using the V0M multiplicity-based estimator~\cite{ALICE:2020nkc}. The \ktopi and \ptopi ratios show an increasing trend going from 50--100\% \orho (low-multiplicity) to 0--1\% \orho (HM) events. This represents a $30\%$ and $27\%$ increase between the two extremes of flattenicity classes for the \ktopi and \ptopi, respectively. In order to compare these results with their multiplicity-dependent counterparts~\cite{ALICE:2020nkc}, the flattenicity-dependent particle ratios are fitted first using the $a-b \times (c-x)^{-1}$ parameterization, where $a,b$, and $c$ are free fit parameters. The fit is then used to quantify the data-to-fit ratio using the flattenicity and multiplicity-dependent measurements, which are shown in the lower panels of Fig.~\ref{fig:integrated_particle_ratios}. The \ktopi measured with the flattenicity selection is marginally higher than the ratio observed in the multiplicity-dependent measurement. However, this is barely significant considering the current uncertainties. By comparing the \Qpp ratios (cf.~Fig.~\ref{fig:spectra_with_rpp}) with the similar ratios computed for the multiplicity-only dependent results measured in V0M event classes~\cite{ALICE:2019dfi,ALICE:2020nkc}, one can observe that the V0M-based event classification produces more pions at low \pt ($\lesssim 500~\mathrm{MeV}/c$). This might result in a larger \ktopi particle ratio when it is measured in flattenicity event classes. This observation is relevant for interpreting the measurements with strange and multi-strange hadrons in the high-multiplicity program at the LHC. A similar effect is seen for the \ptopi ratio measured using the flattenicity estimator, i.e.\ it is above the one measured as a function of multiplicity at high-particle densities, however, these differences are within the systematic uncertainties. The flattenicity-dependent measurements are accompanied by model predictions from \py and \ep. \py predicts no evolution with multiplicity for both particle ratios. On the contrary, \ep describes the multiplicity dependence of the \ktopi ratio, although it underestimates the data. 

\begin{figure}[t]
    \centering
    \hspace{0cm}
    \includegraphics[width=1\textwidth]{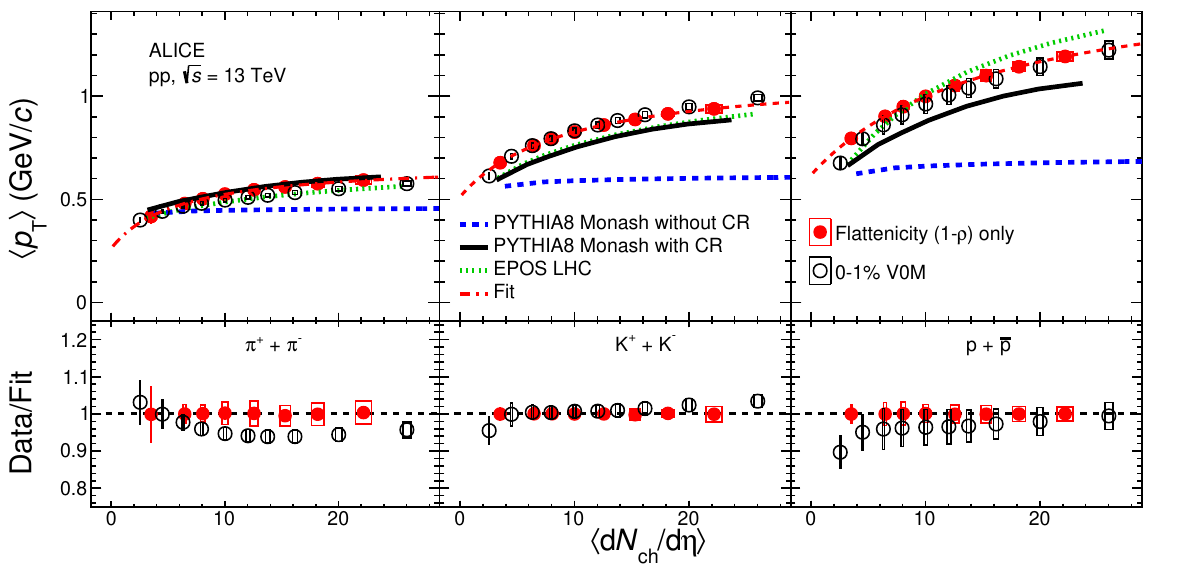}
    \hspace{0cm}
    \caption{Top panels: Average transverse momentum as a function of the average charged-particle density, and predictions using the \py and \ep models. The model calculations and fits (shown with the red dashed lines) correspond to the flattenicity-dependent measurements. The fit uses the $a-b\times(c-x)^{-1}$ parameterization, where $a,b$, and $c$ are free fit parameters. Bottom panels: Data-to-fit ratios for both the flattenicity- and multiplicity-dependent measurements. The multiplicity-dependent results are taken from Ref.~\cite{ALICE:2020nkc}. The statistical and systematic uncertainties are shown with lines and empty boxes.}
    \label{fig:mean_pt}
\end{figure}

\Cref{fig:mean_pt} shows the average transverse momenta of \pion, \kaon, and \pr as a function of the charged-particle density using the flattenicity and V0M multiplicity~\cite{ALICE:2020nkc} based estimators. In both cases, the data show an increasing trend with increasing multiplicity. A mass ordering is observed among the particle species, where protons have the largest \meanpt values. 
The \meanpt of pions with a flattenicity selection is slightly higher than the value observed in the multiplicity-based selection at similar multiplicities. This effect can be attributed to an excess of low-\pt pions $(\lesssim 500~\mathrm{MeV}/c)$ when using the V0M multiplicity estimator~\cite{ALICE:2020nkc}, thereby yielding a lower \meanpt with respect to its counterpart as a function of flattenicity.  
On the other hand, the \meanpt values of kaons and protons are similar within the reported systematic uncertainties between the two selections across the entire multiplicity range. The prediction from \py with color reconnection effects and \ep provide a qualitative description of the data, while \py without color reconnection effects predicts no evolution neither with multiplicity nor with flattenicity. 

\section{Conclusions}\label{sec:Conclusions}

This paper reports on a new event activity estimator named flattenicity (\orho), which can effectively select pp collisions with large number of multiparton interactions with smaller bias due to local multiplicity fluctuations than the multiplicity-based estimator. The local multiplicity fluctuations are due to high-momentum jets affecting the high-\pt particle yield.  To prove this, the transverse momentum spectra of charged pions, kaons, (anti)protons, and unidentified particles are reported as a function of flattenicity and compared with previous multiplicity based results. According to \py, flattenicity is sensitive to multiparton interactions and is less affected by biases toward larger \pt due to local multiplicity fluctuations in the \VZERO acceptance than multiplicity. Therefore, the interpretation of flattenicity is based on the specific implementation of MPIs and the modeling of high-multiplicity pp collisions in PYTHIA. A sample of pp collisions in which multiparton interactions dominate corresponds to the 0--1\% \orho class, whereas a sample of pp collisions with a few MPIs corresponds to the 50--100\% \orho class. The former implicitly includes high-multiplicity pp collisions, and the latter, low-multiplicity pp collisions.
The modification of the \pt distributions as a function of flattenicity with respect to those measured in MB events is quantified by the \Qpp ratio, defined in analogy to the nuclear modification factor widely used in heavy-ion collisions. Models like \py without color reconnection, in which color strings are not allowed to interact with each other, predict a \Qpp close to unity. However, the \Qpp for events in the 0--1\% \orho class exhibits a bump structure at intermediate \pt ($1-8$\,GeV/$c$), while for higher \pt values the \Qpp gradually decreases to the vicinity of unity. The effect is hadron mass dependent. The transverse momentum spectra and \Qpp as a function of \pt for different flattenicity classes are quantitatively described by \py with color reconnection. This observation suggests that pp data cannot be described by a mere superposition of independent parton-parton scatterings. The \ep model overestimates \Qpp at intermediate \pt in particular for the 0--1\% \orho class. To factorize the multiplicity dependence of flattenicity, high multiplicity pp collisions  are analyzed in the same way. Overall, the observations and conclusions are very similar.  
The \pt-integrated particle ratios as a function of flattenicity also exhibit features that have not been observed before. For example, the kaon-to-pion and the proton-to-pion ratios increase from low to high charged-particle multiplicity. Such a increase is slightly steeper than that measured as a function of the V0M multiplicity due to different biases on the multiplicity estimators. These results suggest that flattenicity is a complementary event activity estimator that can help us understand the biases induced when selecting high-multiplicity pp collisions.


\newenvironment{acknowledgement}{\relax}{\relax}
\begin{acknowledgement}
\section*{Acknowledgements}

The ALICE Collaboration would like to thank all its engineers and technicians for their invaluable contributions to the construction of the experiment and the CERN accelerator teams for the outstanding performance of the LHC complex.
The ALICE Collaboration gratefully acknowledges the resources and support provided by all Grid centres and the Worldwide LHC Computing Grid (WLCG) collaboration.
The ALICE Collaboration acknowledges the following funding agencies for their support in building and running the ALICE detector:
A. I. Alikhanyan National Science Laboratory (Yerevan Physics Institute) Foundation (ANSL), State Committee of Science and World Federation of Scientists (WFS), Armenia;
Austrian Academy of Sciences, Austrian Science Fund (FWF): [M 2467-N36] and Nationalstiftung f\"{u}r Forschung, Technologie und Entwicklung, Austria;
Ministry of Communications and High Technologies, National Nuclear Research Center, Azerbaijan;
Conselho Nacional de Desenvolvimento Cient\'{\i}fico e Tecnol\'{o}gico (CNPq), Financiadora de Estudos e Projetos (Finep), Funda\c{c}\~{a}o de Amparo \`{a} Pesquisa do Estado de S\~{a}o Paulo (FAPESP) and Universidade Federal do Rio Grande do Sul (UFRGS), Brazil;
Bulgarian Ministry of Education and Science, within the National Roadmap for Research Infrastructures 2020-2027 (object CERN), Bulgaria;
Ministry of Education of China (MOEC) , Ministry of Science \& Technology of China (MSTC) and National Natural Science Foundation of China (NSFC), China;
Ministry of Science and Education and Croatian Science Foundation, Croatia;
Centro de Aplicaciones Tecnol\'{o}gicas y Desarrollo Nuclear (CEADEN), Cubaenerg\'{\i}a, Cuba;
Ministry of Education, Youth and Sports of the Czech Republic, Czech Republic;
The Danish Council for Independent Research | Natural Sciences, the VILLUM FONDEN and Danish National Research Foundation (DNRF), Denmark;
Helsinki Institute of Physics (HIP), Finland;
Commissariat \`{a} l'Energie Atomique (CEA) and Institut National de Physique Nucl\'{e}aire et de Physique des Particules (IN2P3) and Centre National de la Recherche Scientifique (CNRS), France;
Bundesministerium f\"{u}r Bildung und Forschung (BMBF) and GSI Helmholtzzentrum f\"{u}r Schwerionenforschung GmbH, Germany;
General Secretariat for Research and Technology, Ministry of Education, Research and Religions, Greece;
National Research, Development and Innovation Office, Hungary;
Department of Atomic Energy Government of India (DAE), Department of Science and Technology, Government of India (DST), University Grants Commission, Government of India (UGC) and Council of Scientific and Industrial Research (CSIR), India;
National Research and Innovation Agency - BRIN, Indonesia;
Istituto Nazionale di Fisica Nucleare (INFN), Italy;
Japanese Ministry of Education, Culture, Sports, Science and Technology (MEXT) and Japan Society for the Promotion of Science (JSPS) KAKENHI, Japan;
Consejo Nacional de Ciencia (CONACYT) y Tecnolog\'{i}a, through Fondo de Cooperaci\'{o}n Internacional en Ciencia y Tecnolog\'{i}a (FONCICYT) and Direcci\'{o}n General de Asuntos del Personal Academico (DGAPA), Mexico;
Nederlandse Organisatie voor Wetenschappelijk Onderzoek (NWO), Netherlands;
The Research Council of Norway, Norway;
Pontificia Universidad Cat\'{o}lica del Per\'{u}, Peru;
Ministry of Science and Higher Education, National Science Centre and WUT ID-UB, Poland;
Korea Institute of Science and Technology Information and National Research Foundation of Korea (NRF), Republic of Korea;
Ministry of Education and Scientific Research, Institute of Atomic Physics, Ministry of Research and Innovation and Institute of Atomic Physics and Universitatea Nationala de Stiinta si Tehnologie Politehnica Bucuresti, Romania;
Ministry of Education, Science, Research and Sport of the Slovak Republic, Slovakia;
National Research Foundation of South Africa, South Africa;
Swedish Research Council (VR) and Knut \& Alice Wallenberg Foundation (KAW), Sweden;
European Organization for Nuclear Research, Switzerland;
Suranaree University of Technology (SUT), National Science and Technology Development Agency (NSTDA) and National Science, Research and Innovation Fund (NSRF via PMU-B B05F650021), Thailand;
Turkish Energy, Nuclear and Mineral Research Agency (TENMAK), Turkey;
National Academy of  Sciences of Ukraine, Ukraine;
Science and Technology Facilities Council (STFC), United Kingdom;
National Science Foundation of the United States of America (NSF) and United States Department of Energy, Office of Nuclear Physics (DOE NP), United States of America.
In addition, individual groups or members have received support from:
Czech Science Foundation (grant no. 23-07499S), Czech Republic;
European Research Council (grant no. 950692), European Union;
ICSC - Centro Nazionale di Ricerca in High Performance Computing, Big Data and Quantum Computing, European Union - NextGenerationEU;
Academy of Finland (Center of Excellence in Quark Matter) (grant nos. 346327, 346328), Finland.

\end{acknowledgement}

\bibliographystyle{utphys}   
\bibliography{bibliography}

\appendix
\newpage

\section{The ALICE Collaboration}
\label{app:collab}
\begin{flushleft} 
\small

S.~Acharya\,\orcidlink{0000-0002-9213-5329}\,$^{\rm 127}$, 
D.~Adamov\'{a}\,\orcidlink{0000-0002-0504-7428}\,$^{\rm 86}$, 
A.~Agarwal$^{\rm 135}$, 
G.~Aglieri Rinella\,\orcidlink{0000-0002-9611-3696}\,$^{\rm 32}$, 
L.~Aglietta\,\orcidlink{0009-0003-0763-6802}\,$^{\rm 24}$, 
M.~Agnello\,\orcidlink{0000-0002-0760-5075}\,$^{\rm 29}$, 
N.~Agrawal\,\orcidlink{0000-0003-0348-9836}\,$^{\rm 25}$, 
Z.~Ahammed\,\orcidlink{0000-0001-5241-7412}\,$^{\rm 135}$, 
S.~Ahmad\,\orcidlink{0000-0003-0497-5705}\,$^{\rm 15}$, 
S.U.~Ahn\,\orcidlink{0000-0001-8847-489X}\,$^{\rm 71}$
% ...

\section*{Affiliation Notes}

$^{\rm I}$ Deceased\\
$^{\rm II}$ Also at: Max-Planck-Institut fur Physik, Munich, Germany\\
$^{\rm III}$ Also at: Italian National Agency for New Technologies, Energy and Sustainable Economic Development (ENEA), Bologna, Italy\\
$^{\rm IV}$ Also at: Dipartimento DET del Politecnico di Torino, Turin, Italy\\
$^{\rm V}$ Also at: Yildiz Technical University, Istanbul, T\"{u}rkiye\\
$^{\rm VI}$ Also at: Department of Applied Physics, Aligarh Muslim University, Aligarh, India\\
$^{\rm VII}$ Also at: Institute of Theoretical Physics, University of Wroclaw, Poland\\
$^{\rm VIII}$ Also at: An institution covered by a cooperation agreement with CERN\\

\section*{Collaboration Institutes}

$^{1}$ A.I. Alikhanyan National Science Laboratory (Yerevan Physics Institute) Foundation, Yerevan, Armenia\\
$^{2}$ AGH University of Krakow, Cracow, Poland\\
$^{3}$ Bogolyubov Institute for Theoretical Physics, National Academy of Sciences of Ukraine, Kiev, Ukraine\\
$^{4}$ Bose Institute, Department of Physics  and Centre for Astroparticle Physics and Space Science (CAPSS), Kolkata, India\\
$^{5}$ California Polytechnic State University, San Luis Obispo, California, United States\\
$^{6}$ Central China Normal University, Wuhan, China\\
$^{7}$ Centro de Aplicaciones Tecnol\'{o}gicas y Desarrollo Nuclear (CEADEN), Havana, Cuba\\
$^{8}$ Centro de Investigaci\'{o}n y de Estudios Avanzados (CINVESTAV), Mexico City and M\'{e}rida, Mexico\\
$^{9}$ Chicago State University, Chicago, Illinois, United States\\
$^{10}$ China Institute of Atomic Energy, Beijing, China\\
$^{11}$ China University of Geosciences, Wuhan, China\\
$^{12}$ Chungbuk National University, Cheongju, Republic of Korea\\
$^{13}$ Comenius University Bratislava, Faculty of Mathematics, Physics and Informatics, Bratislava, Slovak Republic\\
$^{14}$ Creighton University, Omaha, Nebraska, United States\\
$^{15}$ Department of Physics, Aligarh Muslim University, Aligarh, India\\
$^{16}$ Department of Physics, Pusan National University, Pusan, Republic of Korea\\
$^{17}$ Department of Physics, Sejong University, Seoul, Republic of Korea\\
$^{18}$ Department of Physics, University of California, Berkeley, California, United States\\
$^{19}$ Department of Physics, University of Oslo, Oslo, Norway\\
$^{20}$ Department of Physics and Technology, University of Bergen, Bergen, Norway\\
$^{21}$ Dipartimento di Fisica, Universit\`{a} di Pavia, Pavia, Italy\\
$^{22}$ Dipartimento di Fisica dell'Universit\`{a} and Sezione INFN, Cagliari, Italy\\
$^{23}$ Dipartimento di Fisica dell'Universit\`{a} and Sezione INFN, Trieste, Italy\\
$^{24}$ Dipartimento di Fisica dell'Universit\`{a} and Sezione INFN, Turin, Italy\\
$^{25}$ Dipartimento di Fisica e Astronomia dell'Universit\`{a} and Sezione INFN, Bologna, Italy\\
$^{26}$ Dipartimento di Fisica e Astronomia dell'Universit\`{a} and Sezione INFN, Catania, Italy\\
$^{27}$ Dipartimento di Fisica e Astronomia dell'Universit\`{a} and Sezione INFN, Padova, Italy\\
$^{28}$ Dipartimento di Fisica `E.R.~Caianiello' dell'Universit\`{a} and Gruppo Collegato INFN, Salerno, Italy\\
$^{29}$ Dipartimento DISAT del Politecnico and Sezione INFN, Turin, Italy\\
$^{30}$ Dipartimento di Scienze MIFT, Universit\`{a} di Messina, Messina, Italy\\
$^{31}$ Dipartimento Interateneo di Fisica `M.~Merlin' and Sezione INFN, Bari, Italy\\
$^{32}$ European Organization for Nuclear Research (CERN), Geneva, Switzerland\\
$^{33}$ Faculty of Electrical Engineering, Mechanical Engineering and Naval Architecture, University of Split, Split, Croatia\\
$^{34}$ Faculty of Engineering and Science, Western Norway University of Applied Sciences, Bergen, Norway\\
$^{35}$ Faculty of Nuclear Sciences and Physical Engineering, Czech Technical University in Prague, Prague, Czech Republic\\
$^{36}$ Faculty of Physics, Sofia University, Sofia, Bulgaria\\
$^{37}$ Faculty of Science, P.J.~\v{S}af\'{a}rik University, Ko\v{s}ice, Slovak Republic\\
$^{38}$ Frankfurt Institute for Advanced Studies, Johann Wolfgang Goethe-Universit\"{a}t Frankfurt, Frankfurt, Germany\\
$^{39}$ Fudan University, Shanghai, China\\
$^{40}$ Gangneung-Wonju National University, Gangneung, Republic of Korea\\
$^{41}$ Gauhati University, Department of Physics, Guwahati, India\\
$^{42}$ Helmholtz-Institut f\"{u}r Strahlen- und Kernphysik, Rheinische Friedrich-Wilhelms-Universit\"{a}t Bonn, Bonn, Germany\\
$^{43}$ Helsinki Institute of Physics (HIP), Helsinki, Finland\\
$^{44}$ High Energy Physics Group,  Universidad Aut\'{o}noma de Puebla, Puebla, Mexico\\
$^{45}$ Horia Hulubei National Institute of Physics and Nuclear Engineering, Bucharest, Romania\\
$^{46}$ HUN-REN Wigner Research Centre for Physics, Budapest, Hungary\\
$^{47}$ Indian Institute of Technology Bombay (IIT), Mumbai, India\\
$^{48}$ Indian Institute of Technology Indore, Indore, India\\
$^{49}$ INFN, Laboratori Nazionali di Frascati, Frascati, Italy\\
$^{50}$ INFN, Sezione di Bari, Bari, Italy\\
$^{51}$ INFN, Sezione di Bologna, Bologna, Italy\\
$^{52}$ INFN, Sezione di Cagliari, Cagliari, Italy\\
$^{53}$ INFN, Sezione di Catania, Catania, Italy\\
$^{54}$ INFN, Sezione di Padova, Padova, Italy\\
$^{55}$ INFN, Sezione di Pavia, Pavia, Italy\\
$^{56}$ INFN, Sezione di Torino, Turin, Italy\\
$^{57}$ INFN, Sezione di Trieste, Trieste, Italy\\
$^{58}$ Inha University, Incheon, Republic of Korea\\
$^{59}$ Institute for Gravitational and Subatomic Physics (GRASP), Utrecht University/Nikhef, Utrecht, Netherlands\\
$^{60}$ Institute of Experimental Physics, Slovak Academy of Sciences, Ko\v{s}ice, Slovak Republic\\
$^{61}$ Institute of Physics, Homi Bhabha National Institute, Bhubaneswar, India\\
$^{62}$ Institute of Physics of the Czech Academy of Sciences, Prague, Czech Republic\\
$^{63}$ Institute of Space Science (ISS), Bucharest, Romania\\
$^{64}$ Institut f\"{u}r Kernphysik, Johann Wolfgang Goethe-Universit\"{a}t Frankfurt, Frankfurt, Germany\\
$^{65}$ Instituto de Ciencias Nucleares, Universidad Nacional Aut\'{o}noma de M\'{e}xico, Mexico City, Mexico\\
$^{66}$ Instituto de F\'{i}sica, Universidade Federal do Rio Grande do Sul (UFRGS), Porto Alegre, Brazil\\
$^{67}$ Instituto de F\'{\i}sica, Universidad Nacional Aut\'{o}noma de M\'{e}xico, Mexico City, Mexico\\
$^{68}$ iThemba LABS, National Research Foundation, Somerset West, South Africa\\
$^{69}$ Jeonbuk National University, Jeonju, Republic of Korea\\
$^{70}$ Johann-Wolfgang-Goethe Universit\"{a}t Frankfurt Institut f\"{u}r Informatik, Fachbereich Informatik und Mathematik, Frankfurt, Germany\\
$^{71}$ Korea Institute of Science and Technology Information, Daejeon, Republic of Korea\\
$^{72}$ KTO Karatay University, Konya, Turkey\\
$^{73}$ Laboratoire de Physique Subatomique et de Cosmologie, Universit\'{e} Grenoble-Alpes, CNRS-IN2P3, Grenoble, France\\
$^{74}$ Lawrence Berkeley National Laboratory, Berkeley, California, United States\\
$^{75}$ Lund University Department of Physics, Division of Particle Physics, Lund, Sweden\\
$^{76}$ Nagasaki Institute of Applied Science, Nagasaki, Japan\\
$^{77}$ Nara Women{'}s University (NWU), Nara, Japan\\
$^{78}$ National and Kapodistrian University of Athens, School of Science, Department of Physics , Athens, Greece\\
$^{79}$ National Centre for Nuclear Research, Warsaw, Poland\\
$^{80}$ National Institute of Science Education and Research, Homi Bhabha National Institute, Jatni, India\\
$^{81}$ National Nuclear Research Center, Baku, Azerbaijan\\
$^{82}$ National Research and Innovation Agency - BRIN, Jakarta, Indonesia\\
$^{83}$ Niels Bohr Institute, University of Copenhagen, Copenhagen, Denmark\\
$^{84}$ Nikhef, National institute for subatomic physics, Amsterdam, Netherlands\\
$^{85}$ Nuclear Physics Group, STFC Daresbury Laboratory, Daresbury, United Kingdom\\
$^{86}$ Nuclear Physics Institute of the Czech Academy of Sciences, Husinec-\v{R}e\v{z}, Czech Republic\\
$^{87}$ Oak Ridge National Laboratory, Oak Ridge, Tennessee, United States\\
$^{88}$ Ohio State University, Columbus, Ohio, United States\\
$^{89}$ Physics department, Faculty of science, University of Zagreb, Zagreb, Croatia\\
$^{90}$ Physics Department, Panjab University, Chandigarh, India\\
$^{91}$ Physics Department, University of Jammu, Jammu, India\\
$^{92}$ Physics Program and International Institute for Sustainability with Knotted Chiral Meta Matter (SKCM2), Hiroshima University, Hiroshima, Japan\\
$^{93}$ Physikalisches Institut, Eberhard-Karls-Universit\"{a}t T\"{u}bingen, T\"{u}bingen, Germany\\
$^{94}$ Physikalisches Institut, Ruprecht-Karls-Universit\"{a}t Heidelberg, Heidelberg, Germany\\
$^{95}$ Physik Department, Technische Universit\"{a}t M\"{u}nchen, Munich, Germany\\
$^{96}$ Politecnico di Bari and Sezione INFN, Bari, Italy\\
$^{97}$ Research Division and ExtreMe Matter Institute EMMI, GSI Helmholtzzentrum f\"ur Schwerionenforschung GmbH, Darmstadt, Germany\\
$^{98}$ Saga University, Saga, Japan\\
$^{99}$ Saha Institute of Nuclear Physics, Homi Bhabha National Institute, Kolkata, India\\
$^{100}$ School of Physics and Astronomy, University of Birmingham, Birmingham, United Kingdom\\
$^{101}$ Secci\'{o}n F\'{\i}sica, Departamento de Ciencias, Pontificia Universidad Cat\'{o}lica del Per\'{u}, Lima, Peru\\
$^{102}$ Stefan Meyer Institut f\"{u}r Subatomare Physik (SMI), Vienna, Austria\\
$^{103}$ SUBATECH, IMT Atlantique, Nantes Universit\'{e}, CNRS-IN2P3, Nantes, France\\
$^{104}$ Sungkyunkwan University, Suwon City, Republic of Korea\\
$^{105}$ Suranaree University of Technology, Nakhon Ratchasima, Thailand\\
$^{106}$ Technical University of Ko\v{s}ice, Ko\v{s}ice, Slovak Republic\\
$^{107}$ The Henryk Niewodniczanski Institute of Nuclear Physics, Polish Academy of Sciences, Cracow, Poland\\
$^{108}$ The University of Texas at Austin, Austin, Texas, United States\\
$^{109}$ Universidad Aut\'{o}noma de Sinaloa, Culiac\'{a}n, Mexico\\
$^{110}$ Universidade de S\~{a}o Paulo (USP), S\~{a}o Paulo, Brazil\\
$^{111}$ Universidade Estadual de Campinas (UNICAMP), Campinas, Brazil\\
$^{112}$ Universidade Federal do ABC, Santo Andre, Brazil\\
$^{113}$ Universitatea Nationala de Stiinta si Tehnologie Politehnica Bucuresti, Bucharest, Romania\\
$^{114}$ University of Cape Town, Cape Town, South Africa\\
$^{115}$ University of Derby, Derby, United Kingdom\\
$^{116}$ University of Houston, Houston, Texas, United States\\
$^{117}$ University of Jyv\"{a}skyl\"{a}, Jyv\"{a}skyl\"{a}, Finland\\
$^{118}$ University of Kansas, Lawrence, Kansas, United States\\
$^{119}$ University of Liverpool, Liverpool, United Kingdom\\
$^{120}$ University of Science and Technology of China, Hefei, China\\
$^{121}$ University of South-Eastern Norway, Kongsberg, Norway\\
$^{122}$ University of Tennessee, Knoxville, Tennessee, United States\\
$^{123}$ University of the Witwatersrand, Johannesburg, South Africa\\
$^{124}$ University of Tokyo, Tokyo, Japan\\
$^{125}$ University of Tsukuba, Tsukuba, Japan\\
$^{126}$ Universit\"{a}t M\"{u}nster, Institut f\"{u}r Kernphysik, M\"{u}nster, Germany\\
$^{127}$ Universit\'{e} Clermont Auvergne, CNRS/IN2P3, LPC, Clermont-Ferrand, France\\
$^{128}$ Universit\'{e} de Lyon, CNRS/IN2P3, Institut de Physique des 2 Infinis de Lyon, Lyon, France\\
$^{129}$ Universit\'{e} de Strasbourg, CNRS, IPHC UMR 7178, F-67000 Strasbourg, France, Strasbourg, France\\
$^{130}$ Universit\'{e} Paris-Saclay, Centre d'Etudes de Saclay (CEA), IRFU, D\'{e}partment de Physique Nucl\'{e}aire (DPhN), Saclay, France\\
$^{131}$ Universit\'{e}  Paris-Saclay, CNRS/IN2P3, IJCLab, Orsay, France\\
$^{132}$ Universit\`{a} degli Studi di Foggia, Foggia, Italy\\
$^{133}$ Universit\`{a} del Piemonte Orientale, Vercelli, Italy\\
$^{134}$ Universit\`{a} di Brescia, Brescia, Italy\\
$^{135}$ Variable Energy Cyclotron Centre, Homi Bhabha National Institute, Kolkata, India\\
$^{136}$ Warsaw University of Technology, Warsaw, Poland\\
$^{137}$ Wayne State University, Detroit, Michigan, United States\\
$^{138}$ Yale University, New Haven, Connecticut, United States\\
$^{139}$ Yonsei University, Seoul, Republic of Korea\\
$^{140}$  Zentrum  f\"{u}r Technologie und Transfer (ZTT), Worms, Germany\\
$^{141}$ Affiliated with an institute covered by a cooperation agreement with CERN\\
$^{142}$ Affiliated with an international laboratory covered by a cooperation agreement with CERN.\\

\end{flushleft} 

\end{document}